%% file: main.tex
\documentclass[12pt,english,twoside]{article}
\usepackage{authblk}

\usepackage{graphicx}
\usepackage{epstopdf,epsfig}
\usepackage[svgnames]{xcolor}
\usepackage{pgfplots}
\usepackage[nomessages]{fp}
\usepackage[np,autolanguage]{numprint}
\npdecimalsign{.}
\pgfplotsset{compat=1.3}
\usetikzlibrary{external}
\usepgfplotslibrary{groupplots}
\tikzset{external/mode=graphics if exists}
\input{lines_symbols}
\input{tikz_options}
\usepgfplotslibrary{colormaps}
\usepackage{amsmath,amssymb,amsfonts,mathrsfs,pictex,color,latexsym}

\usepackage{mathtools}
\usepackage{accents}
\usepackage{gensymb}
\usepackage{textcomp}
\usepackage{siunitx}
\numberwithin{equation}{section} 

\usepackage[colorlinks,linkcolor=blue,citecolor=blue,urlcolor=blue]{hyperref}

\usepackage[
a4paper,
headheight=1.\baselineskip,
top=25mm,
bottom=20mm,
inner=25mm,
outer=25mm,
heightrounded=true]
{geometry}

\usepackage{fancyhdr}
\renewcommand\appendixname{Appendix}

\usepackage{titlesec}

\titleformat{\section}
  {\large\bfseries}
  {\thesection.}
  {0.25em}
  {\vspace{-1ex}}

\titleformat{\subsection}
  {\normalsize\filcenter}
  {\thesubsection.}
  {0.5em}
  {\itshape\vspace{-1ex}}

\usepackage[symbol]{footmisc}
\def\@fnsymbol#1{\ensuremath{\ifcase#1\or \dagger\or \ddagger\or
   \mathsection\or \mathparagraph\or \|\or **\or \dagger\dagger
   \or \ddagger\ddagger \else\@ctrerr\fi}}

\usepackage{array}
\newcolumntype{H}{>{\setbox0=\hbox\bgroup}c<{\egroup}@{}}
\usepackage{arydshln} 
\usepackage[round,sort&compress]{natbib}
\let\doi\undefined
\usepackage{doi}

\usepackage[normalem]{ulem}
\usepackage{etex}
\newcommand{\revisions}[1]{}

\author[1,2]{\bfseries Markus Weyrauch\thanks{Email address for correspondence:
          \href{mailto:markus.scherer@kit.edu}{markus.scherer@kit.edu}}}
\author[2]{\bfseries Moritz Linkmann}
\author[2]{\bfseries Jacob Page}

\affil[1]{\small Institute for Water and Environment -- Numerical Fluid Mechanics Group,
        Karlsruhe Institute of Technology, 76131 Karlsruhe, Germany}
\affil[2]{\small School of Mathematics, University of Edinburgh, Edinburgh, EH9 3FD, UK}

\vspace{-2.2cm}
\date{\raggedright \small (Dated: \today)}
\vspace{-1.2cm}

\title{
  State estimation in homogeneous isotropic turbulence
  using super-resolution with a 4DVar training algorithm
}

\usepackage{titling}
\pretitle{\begin{flushleft}\LARGE\bfseries}
\posttitle{\par\end{flushleft}}
\preauthor{\begin{flushleft}\large\bfseries}
\postauthor{\end{flushleft}}
\predate{\begin{flushleft}}
\postdate{\end{flushleft}}

\input{mathoperators.tex}

\input{nomenclature.tex}

\input{NHS_physparams.tex}

\input{KF_physparams.tex}

\begin{document}

\maketitle
\thispagestyle{fancy}

%
\newcommand{\JacobJFM}{P25} 

\input{sections/abstract}

  \section{Introduction}\label{sec:intro}
  \input{sections/introduction}

  \section{Flow configuration}\label{sec:theory}
  \input{sections/theory}

  \section{Numerical method}\label{sec:numerics}
  \input{sections/numerics}

  \section{Results}\label{sec:results}
  \input{sections/results}

  \section{Conclusion}\label{sec:conclusion}
  \input{sections/conclusion}

  \section*{Funding}
  \input{sections/funding}

  \vspace*{2ex}
  \section*{Declaration of Interests}
  The authors report no conflict of interest.

  \vspace*{2ex}
  \section*{Author ORCID}
  M. Weyrauch,
  \href{https://orcid.org/0000-0002-6301-4704}{https://orcid.org/0000-0002-6301-4704};\\
  M. Linkmann,
  \href{https://orcid.org/0000-0002-3394-1443}{https://orcid.org/0000-0002-3394-1443};\\
  J. Page,
  \href{https://orcid.org/0000-0002-4564-5086}{https://orcid.org/0000-0002-4564-5086}

  \appendix
  \titleformat{\section}
    {\large\bfseries}
    {\appendixname\hspace{0.25em}\thesection.}
    {0.25em}
    {\vspace{-1ex}}

  \section{\DA{} in three-dimensional Kolmogorov flow}\label{sec:appendix3DKolmogorov}
  \input{sections/appendixKF}

  \phantomsection
  \bibliographystyle{plainnatdoilinknourl}
  \setlength{\bibsep}{.4ex}
  \addcontentsline{toc}{section}{References}
  \bibliography{%
    bibForCompile/literature_invsol,%
    bibForCompile/literature_numerics,%
    bibForCompile/literature_dataAssim,%
    bibForCompile/literature_ML,%
    bibForCompile/literature_HIT,%
  }


\end{document}

%% file: lines_symbols.tex
\usepackage{relsize}

\definecolor{my_dark_blue}{rgb}{0.0,0.0,0.5}
\definecolor{my_blue}{rgb}{0.0,0.4,0.8}
\definecolor{my_cyan}{rgb}{0.13,0.82,0.82}
\definecolor{my_lightblue}{rgb}{0.66,0.76,0.83}
\definecolor{my_petrol}{rgb}{0.19,0.55,0.55}
\definecolor{my_green}{rgb}{0.0,0.6,0.51}
\definecolor{my_green2}{rgb}{0.16,0.75,0.47}
\definecolor{my_brgreen}{rgb}{0.67,0.84,0.19}
\definecolor{my_dark_green}{rgb}{0.27,0.54,0.19}
\definecolor{my_orange}{rgb}{0.9,0.55,0.0}
\definecolor{my_red}{rgb}{1.0,0.0,0.0}
\definecolor{my_wine_red}{rgb}{0.83,0.0,0.35}
\definecolor{my_dark_red}{rgb}{0.63,0.12,0.10}
\definecolor{my_purple}{rgb}{0.5,0.0,0.7}
\definecolor{col_qhuntneg}{rgb}{0.1,0.3,0.5}
\definecolor{col_qhuntpos}{rgb}{0.71,0.27,0.15} 
\definecolor{mygray}{rgb}{0.5,0.5,0.5}
\definecolor{col_gray}{rgb}{0.63,0.63,0.63}
\definecolor{col_gray2}{rgb}{0.3,0.3,0.3}
\definecolor{bg-color2}{rgb}{0.156,0.173,0.204}
\colorlet{colDNS}{lightgray}
\colorlet{colInterp}{my_blue}
\colorlet{colSRdyn}{black}
\colorlet{colDA}{my_orange}
\colorlet{colDASR}{my_red}
\colorlet{colGauss}{my_lightblue}

\colorlet{colkf}{darkgray}
\colorlet{colNyquist}{my_purple}
\colorlet{colLalescu}{my_dark_blue}

\definecolor{col1_matlab}{rgb}{0.000,0.447,0.741}
\definecolor{col2_matlab}{rgb}{0.850,0.325,0.098}
\definecolor{col3_matlab}{rgb}{0.929,0.694,0.125}
\definecolor{col4_matlab}{rgb}{0.494,0.184,0.556}
\definecolor{col5_matlab}{rgb}{0.466,0.674,0.188}
\definecolor{col6_matlab}{rgb}{0.301,0.745,0.933}
\definecolor{col7_matlab}{rgb}{0.635,0.078,0.184}

%% file: tikz_options.tex
\pgfplotsset{tikzPlotsDefault/.style=
    {hide axis,scale only axis,
     height=0pt,width=0pt,
     colorbar right,
     colorbar style={
             anchor=west,
             major tick length=0.05cm,
             ylabel near ticks,
             label style={font=\footnotesize},
             ylabel shift={-0.05cm}
             }
     }
}

\pgfplotsset{colorbar style={
        anchor=west,
        major tick length=0.05cm,
        ylabel near ticks,
        label style={font=\small}
        }
}

\pgfplotsset{
    colormap={bluered}{
        rgb255=(0,39,93)
        rgb255=(222,222,222)
        rgb255=(222,100,25)
    },
}
\pgfplotsset{
    colormap={whiteblue}{
        rgb255=(222,222,222)
        rgb255=(  0, 39, 93)
    },
}
\pgfplotsset{
    colormap={whitegreen}{
        rgb255=(250,250,250)
        rgb255=(  0,120, 50)
    },
}
\pgfplotsset{
    colormap={greenwhite}{
        rgb255=(  0,120, 50)
        rgb255=(250,250,250)
    },
}
\pgfplotsset{
    colormap={whitered}{
        rgb255=(222,222,222)
        rgb255=(222, 25, 25)
    },
}
\pgfplotsset{
    colormap={whiteorange}{
        rgb255=(222,222,222)
        rgb255=(229,140,  0)
    },
}
\pgfplotsset{
    colormap={whiteblack}{
        rgb255=(255,255,255)
        rgb255=(  0,  0,  0)
    },
}

\pgfplotsset{
    colormap={bluebrblue}{
    rgb255=(  0, 22,104)
    rgb255=(  3, 25,106)
    rgb255=(  5, 29,109)
    rgb255=(  8, 32,111)
    rgb255=( 11, 35,113)
    rgb255=( 13, 38,116)
    rgb255=( 16, 42,118)
    rgb255=( 19, 45,120)
    rgb255=( 22, 48,123)
    rgb255=( 24, 52,125)
    rgb255=( 27, 55,127)
    rgb255=( 30, 58,130)
    rgb255=( 32, 61,132)
    rgb255=( 35, 65,134)
    rgb255=( 38, 68,137)
    rgb255=( 40, 71,139)
    rgb255=( 43, 74,142)
    rgb255=( 46, 78,144)
    rgb255=( 48, 81,146)
    rgb255=( 51, 84,149)
    rgb255=( 54, 88,151)
    rgb255=( 56, 91,153)
    rgb255=( 59, 94,156)
    rgb255=( 62, 97,158)
    rgb255=( 64,101,160)
    rgb255=( 67,104,163)
    rgb255=( 70,107,165)
    rgb255=( 73,111,167)
    rgb255=( 75,114,170)
    rgb255=( 78,117,172)
    rgb255=( 81,120,174)
    rgb255=( 83,124,177)
    rgb255=( 86,127,179)
    rgb255=( 89,130,181)
    rgb255=( 91,134,184)
    rgb255=( 94,137,186)
    rgb255=( 97,140,188)
    rgb255=( 99,143,191)
    rgb255=(102,147,193)
    rgb255=(105,150,195)
    rgb255=(108,153,198)
    rgb255=(110,157,200)
    rgb255=(113,160,202)
    rgb255=(116,163,205)
    rgb255=(118,166,207)
    rgb255=(121,170,209)
    rgb255=(124,173,212)
    rgb255=(126,176,214)
    rgb255=(129,180,216)
    rgb255=(132,183,219)
    rgb255=(134,186,221)
    rgb255=(137,189,224)
    rgb255=(140,193,226)
    rgb255=(142,196,228)
    rgb255=(145,199,231)
    rgb255=(148,202,233)
    rgb255=(150,206,235)
    rgb255=(153,209,238)
    rgb255=(156,212,240)
    rgb255=(159,216,242)
    rgb255=(161,219,245)
    rgb255=(164,222,247)
    rgb255=(167,225,249)
    rgb255=(169,229,252)
    rgb255=(172,232,254)
    },
}
\pgfplotsset{
    colormap={redorange}{
    rgb255=( 66,  0,  0)
    rgb255=( 69,  0,  0)
    rgb255=( 72,  0,  0)
    rgb255=( 74,  0,  0)
    rgb255=( 77,  0,  0)
    rgb255=( 80,  0,  0)
    rgb255=( 82,  0,  0)
    rgb255=( 85,  0,  0)
    rgb255=( 88,  0,  0)
    rgb255=( 90,  0,  0)
    rgb255=( 93,  0,  0)
    rgb255=( 96,  0,  0)
    rgb255=( 98,  0,  0)
    rgb255=(101,  0,  0)
    rgb255=(104,  0,  0)
    rgb255=(106,  0,  0)
    rgb255=(109,  0,  0)
    rgb255=(112,  0,  0)
    rgb255=(114,  0,  0)
    rgb255=(117,  0,  0)
    rgb255=(120,  0,  0)
    rgb255=(122,  0,  0)
    rgb255=(125,  0,  0)
    rgb255=(128,  0,  0)
    rgb255=(130,  0,  0)
    rgb255=(133,  0,  0)
    rgb255=(135,  0,  0)
    rgb255=(138,  0,  0)
    rgb255=(141,  0,  0)
    rgb255=(143,  0,  0)
    rgb255=(146,  0,  0)
    rgb255=(149,  0,  0)
    rgb255=(151,  0,  0)
    rgb255=(154,  0,  0)
    rgb255=(157,  0,  0)
    rgb255=(159,  0,  0)
    rgb255=(162,  0,  0)
    rgb255=(165,  0,  0)
    rgb255=(167,  0,  0)
    rgb255=(170,  0,  0)
    rgb255=(173,  0,  0)
    rgb255=(175,  0,  0)
    rgb255=(178,  0,  0)
    rgb255=(181,  0,  0)
    rgb255=(183,  0,  0)
    rgb255=(186,  0,  0)
    rgb255=(189,  0,  0)
    rgb255=(191,  0,  0)
    rgb255=(194,  0,  0)
    rgb255=(197,  0,  0)
    rgb255=(199,  0,  0)
    rgb255=(202,  0,  0)
    rgb255=(205,  0,  0)
    rgb255=(207,  0,  0)
    rgb255=(210,  0,  0)
    rgb255=(212,  0,  0)
    rgb255=(215,  0,  0)
    rgb255=(218,  0,  0)
    rgb255=(220,  0,  0)
    rgb255=(223,  0,  0)
    rgb255=(226,  0,  0)
    rgb255=(228,  0,  0)
    rgb255=(231,  0,  0)
    rgb255=(234,  0,  0)
    rgb255=(236,  0,  0)
    rgb255=(239,  0,  0)
    rgb255=(242,  0,  0)
    rgb255=(244,  0,  0)
    rgb255=(247,  0,  0)
    rgb255=(250,  0,  0)
    rgb255=(252,  0,  0)
    rgb255=(255,  0,  0)
    rgb255=(255,  3,  0)
    rgb255=(255,  5,  0)
    rgb255=(255,  8,  0)
    rgb255=(255, 11,  0)
    rgb255=(255, 13,  0)
    rgb255=(255, 16,  0)
    rgb255=(255, 19,  0)
    rgb255=(255, 21,  0)
    rgb255=(255, 24,  0)
    rgb255=(255, 27,  0)
    rgb255=(255, 29,  0)
    rgb255=(255, 32,  0)
    rgb255=(255, 35,  0)
    rgb255=(255, 37,  0)
    rgb255=(255, 40,  0)
    rgb255=(255, 42,  0)
    rgb255=(255, 45,  0)
    rgb255=(255, 48,  0)
    rgb255=(255, 50,  0)
    rgb255=(255, 53,  0)
    rgb255=(255, 56,  0)
    rgb255=(255, 58,  0)
    rgb255=(255, 61,  0)
    rgb255=(255, 64,  0)
    rgb255=(255, 66,  0)
    rgb255=(255, 69,  0)
    rgb255=(255, 72,  0)
    rgb255=(255, 74,  0)
    rgb255=(255, 77,  0)
    rgb255=(255, 80,  0)
    rgb255=(255, 82,  0)
    rgb255=(255, 85,  0)
    rgb255=(255, 88,  0)
    rgb255=(255, 90,  0)
    rgb255=(255, 93,  0)
    rgb255=(255, 96,  0)
    rgb255=(255, 98,  0)
    rgb255=(255,101,  0)
    rgb255=(255,104,  0)
    rgb255=(255,106,  0)
    rgb255=(255,109,  0)
    rgb255=(255,112,  0)
    rgb255=(255,114,  0)
    rgb255=(255,117,  0)
    rgb255=(255,120,  0)
    rgb255=(255,122,  0)
    rgb255=(255,125,  0)
    rgb255=(255,128,  0)
    rgb255=(255,130,  0)
    rgb255=(255,133,  0)
    rgb255=(255,135,  0)
    rgb255=(255,138,  0)
    rgb255=(255,141,  0)
    rgb255=(255,143,  0)
    rgb255=(255,146,  0)
    rgb255=(255,149,  0)
    rgb255=(255,151,  0)
    rgb255=(255,154,  0)
    rgb255=(255,157,  0)
    rgb255=(255,159,  0)
    rgb255=(255,162,  0)
    rgb255=(255,165,  0)
    rgb255=(255,167,  0)
    rgb255=(255,170,  0)
    rgb255=(255,173,  0)
    rgb255=(255,175,  0)
    rgb255=(255,178,  0)
    rgb255=(255,181,  0)
    rgb255=(255,183,  0)
    rgb255=(255,186,  0)
    rgb255=(255,189,  0)
    rgb255=(255,191,  0)
    rgb255=(255,194,  0)
    rgb255=(255,197,  0)
    rgb255=(255,199,  0)
    rgb255=(255,202,  0)
    rgb255=(255,205,  0)
    rgb255=(255,207,  0)
    rgb255=(255,210,  0)
    rgb255=(255,212,  0)
    rgb255=(255,215,  0)
    rgb255=(255,218,  0)
    rgb255=(255,220,  0)
    rgb255=(255,223,  0)
    rgb255=(255,226,  0)
    rgb255=(255,228,  0)
    rgb255=(255,231,  0)
    rgb255=(255,234,  0)
    rgb255=(255,236,  0)
    rgb255=(255,239,  0)
    rgb255=(255,242,  0)
    rgb255=(255,244,  0)
    rgb255=(255,247,  0)
    rgb255=(255,250,  0)
    rgb255=(255,252,  0)
    rgb255=(255,255,  0)
    rgb255=(255,255,  4)
    rgb255=(255,255,  8)
    rgb255=(255,255, 12)
    rgb255=(255,255, 16)
    rgb255=(255,255, 20)
    rgb255=(255,255, 24)
    rgb255=(255,255, 28)
    rgb255=(255,255, 32)
    rgb255=(255,255, 36)
    rgb255=(255,255, 40)
    rgb255=(255,255, 44)
    rgb255=(255,255, 48)
    rgb255=(255,255, 52)
    rgb255=(255,255, 56)
    rgb255=(255,255, 60)
    rgb255=(255,255, 64)
    rgb255=(255,255, 68)
    rgb255=(255,255, 72)
    rgb255=(255,255, 76)
    rgb255=(255,255, 80)
    rgb255=(255,255, 84)
    rgb255=(255,255, 88)
    rgb255=(255,255, 92)
    rgb255=(255,255, 96)
    rgb255=(255,255,100)
    rgb255=(255,255,104)
    rgb255=(255,255,108)
    rgb255=(255,255,112)
    rgb255=(255,255,116)
    },
}
\pgfplotsset{
colormap={bluebrblueorangered}{
    rgb255=(    0,  22, 104)
    rgb255=(    3,  25, 106)
    rgb255=(    5,  29, 109)
    rgb255=(    8,  32, 111)
    rgb255=(   11,  35, 113)
    rgb255=(   13,  38, 116)
    rgb255=(   16,  42, 118)
    rgb255=(   19,  45, 120)
    rgb255=(   22,  48, 123)
    rgb255=(   24,  52, 125)
    rgb255=(   27,  55, 127)
    rgb255=(   30,  58, 130)
    rgb255=(   32,  61, 132)
    rgb255=(   35,  65, 134)
    rgb255=(   38,  68, 137)
    rgb255=(   40,  71, 139)
    rgb255=(   43,  74, 142)
    rgb255=(   46,  78, 144)
    rgb255=(   48,  81, 146)
    rgb255=(   51,  84, 149)
    rgb255=(   54,  88, 151)
    rgb255=(   56,  91, 153)
    rgb255=(   59,  94, 156)
    rgb255=(   62,  97, 158)
    rgb255=(   64, 101, 160)
    rgb255=(   67, 104, 163)
    rgb255=(   70, 107, 165)
    rgb255=(   73, 111, 167)
    rgb255=(   75, 114, 170)
    rgb255=(   78, 117, 172)
    rgb255=(   81, 120, 174)
    rgb255=(   83, 124, 177)
    rgb255=(   86, 127, 179)
    rgb255=(   89, 130, 181)
    rgb255=(   91, 134, 184)
    rgb255=(   94, 137, 186)
    rgb255=(   97, 140, 188)
    rgb255=(   99, 143, 191)
    rgb255=(  102, 147, 193)
    rgb255=(  105, 150, 195)
    rgb255=(  108, 153, 198)
    rgb255=(  110, 157, 200)
    rgb255=(  113, 160, 202)
    rgb255=(  116, 163, 205)
    rgb255=(  118, 166, 207)
    rgb255=(  121, 170, 209)
    rgb255=(  124, 173, 212)
    rgb255=(  126, 176, 214)
    rgb255=(  129, 180, 216)
    rgb255=(  132, 183, 219)
    rgb255=(  134, 186, 221)
    rgb255=(  137, 189, 224)
    rgb255=(  140, 193, 226)
    rgb255=(  142, 196, 228)
    rgb255=(  145, 199, 231)
    rgb255=(  148, 202, 233)
    rgb255=(  150, 206, 235)
    rgb255=(  153, 209, 238)
    rgb255=(  156, 212, 240)
    rgb255=(  159, 216, 242)
    rgb255=(  161, 219, 245)
    rgb255=(  164, 222, 247)
    rgb255=(  167, 225, 249)
    rgb255=(  169, 229, 252)
    rgb255=(  172, 232, 254)
    rgb255=(  255, 255, 255)
    rgb255=(  255, 255, 108)
    rgb255=(  255, 255,  96)
    rgb255=(  255, 255,  84)
    rgb255=(  255, 255,  72)
    rgb255=(  255, 255,  60)
    rgb255=(  255, 255,  48)
    rgb255=(  255, 255,  36)
    rgb255=(  255, 255,  24)
    rgb255=(  255, 255,  12)
    rgb255=(  255, 255,   0)
    rgb255=(  255, 247,   0)
    rgb255=(  255, 239,   0)
    rgb255=(  255, 231,   0)
    rgb255=(  255, 223,   0)
    rgb255=(  255, 215,   0)
    rgb255=(  255, 207,   0)
    rgb255=(  255, 199,   0)
    rgb255=(  255, 191,   0)
    rgb255=(  255, 183,   0)
    rgb255=(  255, 175,   0)
    rgb255=(  255, 167,   0)
    rgb255=(  255, 159,   0)
    rgb255=(  255, 151,   0)
    rgb255=(  255, 143,   0)
    rgb255=(  255, 135,   0)
    rgb255=(  255, 128,   0)
    rgb255=(  255, 120,   0)
    rgb255=(  255, 112,   0)
    rgb255=(  255, 104,   0)
    rgb255=(  255,  96,   0)
    rgb255=(  255,  88,   0)
    rgb255=(  255,  80,   0)
    rgb255=(  255,  72,   0)
    rgb255=(  255,  64,   0)
    rgb255=(  255,  56,   0)
    rgb255=(  255,  48,   0)
    rgb255=(  255,  40,   0)
    rgb255=(  255,  32,   0)
    rgb255=(  255,  24,   0)
    rgb255=(  255,  16,   0)
    rgb255=(  255,   8,   0)
    rgb255=(  255,   0,   0)
    rgb255=(  247,   0,   0)
    rgb255=(  239,   0,   0)
    rgb255=(  231,   0,   0)
    rgb255=(  223,   0,   0)
    rgb255=(  215,   0,   0)
    rgb255=(  207,   0,   0)
    rgb255=(  199,   0,   0)
    rgb255=(  191,   0,   0)
    rgb255=(  183,   0,   0)
    rgb255=(  175,   0,   0)
    rgb255=(  167,   0,   0)
    rgb255=(  159,   0,   0)
    rgb255=(  151,   0,   0)
    rgb255=(  143,   0,   0)
    rgb255=(  135,   0,   0)
    rgb255=(  128,   0,   0)
    rgb255=(  120,   0,   0)
    rgb255=(  112,   0,   0)
    rgb255=(  104,   0,   0)
    rgb255=(   96,   0,   0)
    rgb255=(   88,   0,   0)
    rgb255=(   80,   0,   0)
    rgb255=(   72,   0,   0)
    },
}

%% file: mathoperators.tex
\newcommand\abs[1]{\ensuremath{\vert {#1} \vert}}
\newcommand\norm[1]{\ensuremath{\| {#1} \|}}

%% file: nomenclature.tex
\newcommand\interpop{\ensuremath{\mathcal{I}}}
\newcommand\coarseop{\ensuremath{\mathcal{C}}}
\newcommand\observop{\ensuremath{\mathcal{H}}}
\newcommand\nnetweigts{\ensuremath{{\boldsymbol{\theta}}}}
\newcommand\nnetop{\ensuremath{\mathcal{N}_\nnetweigts}}
\newcommand\Lerayop{\ensuremath{\mathcal{L}_{\boldsymbol{u}}}}

\newcommand\DNS{DNS}

\newcommand\SRdyn{SRdyn}
\newcommand\DA{4DVar}
\newcommand\DASR{4DVarSR}

\newcommand\LossfctSR{\ensuremath{\mathscr{L}_{SR}}}

\newcommand\LossfctTC{\ensuremath{\mathscr{L}_{TC}}}
\newcommand\LossfctInt{\ensuremath{\mathscr{L}_{IC}}}
\newcommand\LossfctDA{\ensuremath{\mathscr{L}_{DA}}}

\newcommand\obsspace{\ensuremath{\mathscr{O}}}
\newcommand\physspace{\ensuremath{\mathscr{P}}}
\newcommand\NS{\ensuremath{N_S}}
\newcommand\NT{\ensuremath{N_T}}

\newcommand\Ntraj{\ensuremath{N_{traj}}}
\newcommand\nc{\ensuremath{M}} 
\newcommand\nct{\ensuremath{M_T}} 
\newcommand\Tunroll{\ensuremath{T_u}} %
\newcommand\TunrollDA{\ensuremath{T_{DA}}} %
\newcommand\lrate{\ensuremath{\eta}} %

\newcommand\tfm[1]{\ensuremath{\boldsymbol{\varphi}_{#1}}}

\newcommand\Ekspec{\ensuremath{E}}

\newcommand\Pkspec{\ensuremath{\Pi}}
\newcommand\Crosscorr{\ensuremath{\widehat{\rho}}_{\uvec\uvecpred}}

\newcommand\kc{\ensuremath{k_C}}
\newcommand\lc{\ensuremath{l_C}}
\newcommand\kNyquist{\ensuremath{k_N}}

\newcommand\erruvw{\ensuremath{\epsilon_{\boldsymbol{u}}}}

\newcommand\avec{\ensuremath{\boldsymbol{a}}}

\newcommand\fvecext{\ensuremath{\boldsymbol{f_e}}}
\newcommand\Hvec{\ensuremath{\boldsymbol{H}}}
\newcommand\hvec{\ensuremath{\boldsymbol{h}}}
\newcommand\kvec{\ensuremath{\boldsymbol{k}}}

\newcommand\qvec{\ensuremath{\boldsymbol{q}}}

\newcommand\vvec{\ensuremath{\boldsymbol{v}}}

\newcommand\xvec{\ensuremath{\boldsymbol{x}}}

\newcommand\yvec{\ensuremath{\boldsymbol{y}}}

\newcommand\nablavec{\ensuremath{\boldsymbol{\nabla}}}
\newcommand\Regen{\ensuremath{Re}}    
\newcommand\Relambda{\ensuremath{Re_{\lambda}}}    
\newcommand\ReLeddy{\ensuremath{Re_{L}}}    
\newcommand\lambdaTaylor{\ensuremath{\lambda_T}}    
\newcommand\Leddy{\ensuremath{L_e}}    

\newcommand\forceampl{\ensuremath{\chi}}

\newcommand\kf{\ensuremath{k_f}}

\newcommand\Ekin{\ensuremath{\bar{E}}}

\newcommand\eps{\ensuremath{\varepsilon}}
\newcommand\Diss{\ensuremath{\overline{\varepsilon}}}

\newcommand\realn{\ensuremath{\mathbb{R}}}

\newcommand\dom{\ensuremath{\Omega}}

\newcommand\teddy{\ensuremath{T_e}}
\newcommand\teddyLi{\ensuremath{T_{f}}}

\newcommand\fnu{\ensuremath{\nu}} 
\newcommand\leta{\ensuremath{\eta_K}} 

\newcommand\kxyz{\ensuremath{\boldsymbol{k}}}

\newcommand\kmax{\ensuremath{k_{m}}}

\newcommand\Lxyz{\ensuremath{{L}}}

\newcommand\uvec{\ensuremath{\boldsymbol{u}}}
\newcommand\uveczero{\ensuremath{\boldsymbol{u}_{0}}}

\newcommand\uvecpred{\ensuremath{\boldsymbol{u}^{p}}}
\newcommand\uvecpredzero{\ensuremath{\boldsymbol{u}^{p}_{0}}}

\newcommand\uvecj{\ensuremath{\boldsymbol{u}_j}}

\newcommand\omvec{\ensuremath{\boldsymbol{\omega}}}
\newcommand\omvecpred{\ensuremath{\boldsymbol{\omega}^p}}
\newcommand\omvecpredzero{\ensuremath{\boldsymbol{\omega}^p_0}}
\newcommand\omx{\ensuremath{\omega_{x}}}
\newcommand\omy{\ensuremath{\omega_{y}}}
\newcommand\omz{\ensuremath{\omega_{z}}}

\newcommand\omypred{\ensuremath{\omega_{y}^{p}}}

\newcommand\urms{\ensuremath{u_{rms}}}
\newcommand\uprime{\ensuremath{u^{\prime}}}

\newcommand\omrms{\ensuremath{\omega_{rms}}}
\newcommand\omprime{\ensuremath{\omega^{\prime}}}

\newcommand\uvecfour{\ensuremath{\hat{\uvec}}}

\newcommand\uvecpredfour{\ensuremath{\hat{\uvec}^{p}}}

\newcommand\Nx{\ensuremath{N_x}}
\newcommand\Ny{\ensuremath{N_y}}
\newcommand\Nz{\ensuremath{N_z}}
\newcommand\Ni{\ensuremath{N_i}}
\newcommand\deltatc{\ensuremath{\Delta t_c}} 
\newcommand\deltat{\ensuremath{\Delta t}} 
\newcommand\deltax{\ensuremath{\Delta x}}

%% file: NHS_physparams.tex
\def\NHSeta{0.036105143316090105}

\def\NHSteddy{1.1390660119056701}

\def\NHSLeddy{4.848941855430603}
\def\NHSlambdaTaylor{0.5840020698308944}

\def\NHSReLeddy{689.0398443603516}
\def\NHSRelambda{67.61577590942383}

%% file: KF_physparams.tex
\def\KFeta{0.041841548569500446}

\def\KFLeddy{6.634440712928772}
\def\KFlambdaTaylor{0.7133622843027115}

\def\KFReLeddy{862.3410144042969}
\def\KFRelambda{75.22830986022949}

%% file: sections/abstract.tex
\noindent
Variational data assimilation and machine-learning based super-resolution are two alternative approaches to state estimation in turbulent flows. The former is an optimisation problem featuring a time series of coarse observations, the latter usually requires a library of high-resolution `ground truth’ data. We show that the classic `4DVar’ data assimilation algorithm can be used to train neural networks for super-resolution in three-dimensional isotropic turbulence without the need for high-resolution reference data. To do this, we adapt a pseudo-spectral version of the fully-differentiable JAX-CFD solver (Kochkov et al, \emph{Proc. Nat. Acad. Sci.} \textbf{118}, 2021) to three-dimensional flows and combine it with a convolutional neural network for super-resolution. As a result we are able to include entire trajectories in our loss function which is minimised with gradient-based optimisation to define the neural network weights.
We show that the resulting neural networks outperform 4DVar for state estimation at initial time over a wide variety of metrics, though 4DVar leads to more robust predictions towards the end of its assimilation window. We also present a hybrid approach in which the trained neural network output is used to initialise 4DVar. The resulting performance is more than twice as accurate as other state estimation strategies for all times and performs well even beyond known limiting lengthscales, all without requiring access to high-resolution measurements at any point.

%% file: sections/introduction.tex
Chaotic dynamical systems are notoriously sensitive to small variations in the initial conditions.
Numerical forecasts initiated with an imperfect representation of the current system state will therefore deviate from the evolution of the true system, regardless of how accurately the numerical model captures the governing physics.
A popular family of methods to account for these deviations in numerical forecasting techniques is \emph{data assimilation} (DA), which augments the numerical prediction in regular intervals with observations of the true system state to ensure that the prediction stays in the neighbourhood of the true system trajectory \citep{Talagrand_1997}.
These state estimation techniques usually rely on comparison of the forward propagation of predictions in time, with comparison to observational data.
They are complemented by recent alternatives to the state estimation problem which utilise deep neural networks.
This approach, usually termed `super-resolution’ \citep[see e.g.][]{Fukami_Fukagata_Taira_2023}, can perform well with limited observations, but usually relies on the existence of a large dataset of high-resolution snapshots for network training.
This paper seeks to combine both approaches in a three-dimensional turbulent flow, infusing the training process of a neural network with a data-assimilation algorithm (which does not rely on high-resolution reference data) and initialising assimilation computations with the output of a neural network.

The current state of the art in data assimilation are variational methods, ensemble Kalman filters or combinations of  both \citep{Bannister_2017}.
Classical Kalman filters \citep{Kalman_1960} are sequential techniques that first advance a system state in time until observations become available.
New measurements are used to improve the prediction, with individual observations weighted depending on their uncertainty \citep{Evensen_2007}.
For large-scale non-linear systems, ensemble Kalman filters replace a single trajectory with an ensemble of initial conditions and update the prediction based on
statistics of the ensemble \citep{Evensen_1994}.

%
Variational techniques like \DA{} (4D = three spatial dimensions and time) solve a constrained optimisation problem to determine the initial condition that minimises the deviation between observations and predictions over an assimilation window $0 \leq t \leq \TunrollDA$ in a least squares sense \citep{Talagrand_Courtier_1987}.
Gradients of the predictive model with respect to the initial condition are usually computed via adjoint methods \citep{Kalnay_2002}.
An alternative to coding and backward-time marching an adjoint system is the use of a fully-differentiable solver.
In this approach, gradients are computed via automatic differentiation through the computational graph.
Automatic differentiation back-propagates gradients through the forward computation using a combination of the chain rule (which requires storage of the forward state at each node in the graph) with known derivatives of elementary mathematical operations \citep{Baydin_Pearlmutter_Radul_Siskind_2018}. As a result, gradients are accurate to machine precision.

In the past, both \DA{}
\citep{Bewley_Protas_2004,Gronskis_Heitz_Memin_2013,Wang_Wang_Zaki_2019,%
Li_Zhang_Dong_Abdullah_2020,Wang_Wang_Zaki_2022}
and (ensemble) Kalman filters
\citep{Hoepffner_Chevalier_Bewley_Henningson_2005,Chevalier_Hoepffner_Bewley_Henningson_2006,%
Colburn_Cessna_Bewley_2011,Kato_Obayashi_2013,Kato_Yoshizawa_Ueno_Obayashi_2015}
have been applied in a variety of transitional and fully-developed turbulent flows, with a strong focus on wall-bounded configurations \citep{Buchta_Zaki_2021,Wang_Wang_Zaki_2019,Wang_Zaki_2021,Wang_Wang_Zaki_2022,Wang_Zaki_2025};
see \citealp{Hayase_2015}, \citealp{Mons_Chassaing_Sagaut_2017} and \citealp{Zaki_2025} for a comprehensive overview.
We consider here the state estimation problem in triply-periodic, body-forced turbulence. To our knowledge, 4DVar has been considered in this configuration only by \citet{Li_Zhang_Dong_Abdullah_2020}, who drove the flow with a monochromatic, unidirectional body force.
%
In \citet{Li_Zhang_Dong_Abdullah_2020}, the observations available to the \DA{} algorithm are low-pass filtered velocity snapshots from a target trajectory.
State estimation is robust when the cut-off wavenumber of the spectral filter is not significantly smaller than a critical value $\kc\approx0.2\leta^{-1}$ (equivalent to a critical length scale $\lc\approx5\pi\leta$, with $\leta$ denoting the Kolmogorov length).
The fact that \DA{} struggles as the observational data is coarsened in space is consistent with earlier findings on turbulence synchronisation.
Numerical studies of chaos synchronisation in turbulent flows have found that the small scales (wavenumbers $k>k_C$) of a turbulent flow can be reconstructed by timestepping high-pass-filtered Navier-Stokes equations if all modes up to a wavenumber $k\leq\kc$ are prescribed
\citep{Yoshida_Yamaguchi_Kaneda_2005,Lalescu_Meneveau_Eyink_2013,VelaMartin_2021,Inubushi_Caulfield_2025}.
The indication is that the small-scale dynamics, $k> k_C$, is slaved to the large-scale evolution.
Note that a slightly different problem has also been considered in a channel \citep{Wang_Zaki_2022}, where synchronisation of masked vertical slabs occurs if the height of the masked region is less than the local Taylor microscale.

With the increasing popularity of machine learning techniques
\citep{Brunton_Noack_Koumoutsakos_2020,Brenner_Eldredge_Freund_2019,Taira_Rigas_Fukami_2025},
a variety of data-driven alternatives for state estimation have emerged, though connections with classical state estimation methods are not usually emphasised.
A popular example are super-resolution (SR) methods that were originally developed for image processing.
In these approaches, a model, typically a convolutional neural network (CNN), is trained to reconstruct high-resolution images based on low-quality inputs.
This is accomplished by training the model to minimise a reconstruction error on a library of high-resolution target images \citep{Dong_Loy_He_Tang_2016}.
These methods are straightforward to adapt to turbulent flows, where the goal is the reconstruction of high-resolution velocity fields from coarse observations.
The networks learn data-driven interpolation operators whose ability to reconstruct high-resolution fields outperforms standard polynomial interpolation \citep{Fukami_Fukagata_Taira_2019,Fukami_Fukagata_Taira_2021,Kim_Kim_Won_Lee_2021}.
Recently, more advanced SR techniques have incorporated additional terms in the training loss functions to encourage the reconstructions to satisfy physical constraints (e.g. mass conservation) or to improve reconstruction of velocity gradients/vorticity; a detailed review of different approaches and model architectures can be found in \citet{Fukami_Fukagata_Taira_2023}.

We view super-resolution and data assimilation techniques as complementary approaches to state estimation:
While \DA{} estimates the state given observations of a \emph{single} trajectory,
SR models see a wide variety of flow states sampled from the turbulent attractor but have \emph{no} knowledge of the dynamics along any specific trajectory.
Recent work in data assimilation hints at the potential for improving classical assimilation techniques with the incorporation of super-resolution methods.
Some examples include neural networks that increase the flow field resolution before the assimilation step in ensemble Kalman filters \citep{Barthelemy_Brajard_Bertino_Counillon_2022} or techniques in which data assimilation and super-resolution are performed simultaneously by a neural network \citep{Yasuda_Onishi_2023}.
Another promising approach is to generate high-quality initial guesses for \DA{} with the aid of super-resolution as in \citet{Frerix_al_2021}, where a CNN learned an approximate inverse of the observation operator that maps physical data onto the observation space.

A caveat most of these hybrid techniques share with pure SR methods is a continued reliance on high-resolution training data, which limits their applicability in situations where large, fully-resolved datasets for training are not available (e.g. experimental measurements).
Interestingly, recent observations indicate that models trained exclusively on low-resolution data can estimate high-resolution fields when physical principles are incorporated in the training process.
For example, \citet{Kelshaw_Rigas_Magri_2022} trained such a `physics-informed' network on low-Reynolds number ($Re$) 2D Kolmgorov flow with a loss function that includes a penalisation when the network output does not satisfy the Navier-Stokes equations at each node of the coarse grid. 
The same group have also recently presented a method to generate three-dimensional flow fields from planar observations in a low-$Re$, chaotic Kolmogorov flow \citep{Mo_Magri_2025}.
In another approach, \citet{Page_2025} (hereafter \JacobJFM{}) designed a loss function in analogy to the classical \DA{} optimisation problem by including a numerical time marching scheme in the training process
(`solver-in-the-loop', \citealp{Um_al_2020}).
In contrast to standard SR training on individual high-resolution fields, the super-resolved low-resolution snapshots are time marched using a fully-differentiable DNS code.
This trajectory in then coarse-grained for comparison to the reference coarse observations, just as in classical \DA{}.
The method was demonstrated in two-dimensional Kolmogorov turbulence at fairly high $Re$, where performance of the model was found to be comparable to high-resolution-based super-resolution approaches.
It also outperforms classical 4DVar at initial state estimation for all the levels of coarse-graining that were tested, though the model predictions deviate much faster from the reference trajectory than the 4DVar approach once unrolled in time.
In this paper we consider the state-estimation problem in three-dimensional, homogeneous, isotropic turbulence, using combinations of neural network super-resolution and 4DVar -- none of which require reference high-resolution data.
We first adapt the 4DVar-inspired super-resolution approach of \JacobJFM{} to the 3D problem and assess its capabilities.
To do this, we have developed a spectral, fully-differentiable 3D Navier-Stokes solver around the JAX-CFD data structures introduced in the landmark work by \citet{Kochkov_al_2021} and \citet{Dresdner_al_2022}.
This allows us to include entire trajectories of the 3D flow in our loss function for network training, and also to perform classical 4DVar using automatic differentiation.
We explore the limits of the super-resolution based approach with respect to known limiting length scales, and also demonstrate that 4DVar calculations initialised with super-resolved outputs
performs more than two times better than other methods for state reconstruction
over the entire assimilation window and beyond.

The remainder of this manuscript is organised as follows: In \S\ref{sec:theory}, the flow configuration is presented and relevant physical measures are introduced.
The essentials of the differentiable DNS solver are summarised in \S\ref{sec:numerics}, alongside an overview of the data assimilation and super-resolution techniques used in this study.
In \S\ref{sec:results}, the predictive performance of our trajectory-based super-resolution technique is compared to other state estimation methods and the potential of using the super-resolved fields for the initialization of \DA{} is assessed.
The manuscript closes with a summary of the relevant findings and an outlook on
future work in \S\ref{sec:conclusion}.

%% file: sections/theory.tex
%
We consider the non-dimensional Navier-Stokes equations for an incompressible Newtonian fluid under the action of an external body force $\fvecext$
\begin{alignat}{2}
  \partial_{t}\,\uvec
  + (\uvec \boldsymbol{\cdot} \nablavec) \uvec
  &= - \nablavec p + \dfrac{1}{\Regen} \Delta \uvec + \fvecext,
    \label{eq:theory_NavierStokes_dim}\\[0.5ex]
  \nablavec \boldsymbol{\cdot} \uvec
  &=\;0,
    \label{eq:theory_continuity_dim}
\end{alignat}
in a triply-periodic domain $\dom=[0,L)^3$. Velocity, vorticity and pressure fields are denoted as $\uvec=(u,v,w)^T$, $\omvec=\nablavec\times\uvec=(\omx,\omy,\omz)^T$ and $p$, respectively.
Equation~\eqref{eq:theory_NavierStokes_dim} has been non-dimensionalised by the characteristic length and time scales $L/2\pi$ and $\sqrt{L/(2\pi\forceampl)}$, respectively, leading to a Reynolds number $\Regen = (\sqrt{\forceampl}/\fnu)(L/2\pi)^{3/2}$. Here, $\fnu$ and $\forceampl$ denote the kinematic viscosity and a representative forcing amplitude per unit mass (forcing profile defined below), respectively.
A deterministic, isotropic and non-helical forcing is chosen to sustain turbulent fluctuations
by continuously injecting energy into a range of low wavenumbers \citep{McKay_Linkmann_Clark_Chalupa_Berera_2017, Linkmann2018}. In dimensional form the forcing is
\begin{equation}
  \fvecext^*= \chi \displaystyle\sum_{k=1}^{\kf}
  \left(
  \begin{array}{c}
    \sin(2\pi k z^*/L) + \sin(2\pi k y^*/L) \\
    \sin(2\pi k x^*/L) + \sin(2\pi k z^*/L) \\
    \sin(2\pi k y^*/L) + \sin(2\pi k x^*/L)
  \end{array}
  \right).
\end{equation}

In order to assess the ability of the various state-estimation techniques to accurately reproduce turbulent snapshots, we will compare various spectral quantities which we introduce here for convenience.
We focus on standard observables assessing the energy content at a given scale, that is, the kinetic energy spectrum
\begin{equation}
   \Ekspec(k,t) = \frac{1}{2}\int_{\abs{\kxyz}=k} d \kxyz \ |\uvecfour(\kvec,t)|^2 \  ,
\end{equation}
where $\uvecfour$ is the Fourier transform of the velocity field, and the mean flux of kinetic energy across scales
\begin{equation}
\Pkspec(k,t) = -\langle \tilde{S}^k_{ij} (\widetilde{u_iu_j}^k - \tilde{u}^k_i\tilde{u}^k_j)  \rangle_\Omega \ ,
\end{equation}
where $\tilde{(\cdot)}^k$ refers to Galerkin truncation at wavenumber $k$ and $S_{ij}=(\partial_i u_j + \partial_j u_i)/2$ is the rate-of-strain tensor. The angled brackets denote a spatial average over the domain $\Omega$.

Throughout this paper, time and length scales will be measured in terms of the large-eddy time $\teddy = \urms^2/\Diss$ and the Kolmogorov length $\leta = (\fnu^3/\Diss)^{1/4}$, respectively, where $\urms = \sqrt{2 \Ekin /3}$  is the root-mean-square velocity calculated from the spatio-temporally averaged kinetic energy, $\Ekin$, in a statistically steady state, and $\Diss$ is the spatio-temporal average of the dissipation rate $\varepsilon(\xvec,t) = 2 \nu S_{ij}S_{ji}$.
The Taylor- and large-eddy Reynolds numbers are defined as
$\Relambda = \urms\lambdaTaylor/\fnu$ and
$\ReLeddy = \urms\Leddy/\fnu = (3/20)\Relambda^2$, respectively, where
$\lambdaTaylor=\sqrt{15\urms^2\fnu/\Diss}$ is the Taylor microscale
and $\Leddy=(\Ekin)^{3/2}/\Diss$ is a length scale associated with the large eddies \citep{Pope_2000}.

%
%
\input{sections/tables/parameter_tables.tex}
%
\nprounddigits{0}
\FPeval\TtotPerTraj{round(100/\NHSteddy:0)}
\autoref{tab:param_phys_num} summarises the relevant physical and numerical parameters of the simulations that will be analysed in the remainder of this study.
The values are obtained from ensemble averages over $100$ independent trajectories each of length $\TtotPerTraj\teddy$ (equivalent to $100$ non-dimensional time units).
At the chosen parameter point, the flow attains a Taylor Reynolds number of $\Relambda\approx 70$ (equivalent to $\Regen = 50$ and $\ReLeddy\approx 690$), leading to a separation between the system's largest and smallest scales of $\Leddy/\leta\approx\np{\Leddyoeta}$.
These values are comparable to \citet{Li_Zhang_Dong_Abdullah_2020} who had $\Relambda \approx 75$, albeit with a different forcing profile (see appendix \ref{sec:appendix3DKolmogorov} for an assessment of \DA{} in their flow).
\nprounddigits{2}

%% file: sections/tables/parameter_tables.tex
\FPeval\Lbox{2*\FPpi}
\FPeval\LeddyoL{\NHSLeddy/\Lbox}
\FPeval\lambdaoL{\NHSlambdaTaylor/\Lbox}
\FPeval\Loleta{\Lbox/\NHSeta}
\FPeval\lambdaoleta{\NHSlambdaTaylor/\NHSeta}
\FPeval\Leddyoeta{\NHSLeddy/\NHSeta}

\FPeval\kfEta{3 * \NHSeta}
\FPeval\kmaxEta{trunc(trunc(2./3.*128:0)/2.:0) * \NHSeta}
\FPeval\kNyquistetaOne{128/(2*4)*\NHSeta}
\FPeval\kNyquistetaTwo{128/(2*8)*\NHSeta}
\FPeval\kNyquistetaThree{128/(2*16)*\NHSeta}
\FPeval\dxEta{\Lbox/128 / \NHSeta}
\FPeval\dxcEtaOne{\Lbox/128 * 4 / \NHSeta}
\FPeval\dxcEtaTwo{\Lbox/128 * 8 / \NHSeta}
\FPeval\dxcEtaThree{\Lbox/128 * 16 / \NHSeta}
\FPeval\dxlc{\dxEta/(5*\FPpi)}
\FPeval\dxlcOne{\dxcEtaOne/(5*\FPpi)}
\FPeval\dxlcTwo{\dxcEtaTwo/(5*\FPpi)}
\FPeval\dxlcThree{\dxcEtaThree/(5*\FPpi)}
\FPeval\lcdx{1/\dxlc}

\nprounddigits{2}

\begin{table}
  \begin{center}
    \hrulefill \\
    \hfill \\
    \begin{tabular}{c c H S S S S S S S S}
      \multicolumn{1}{c}{Setting}&
      \multicolumn{1}{c}{$\Ni$}&
      \multicolumn{1}{H}{$\kf\leta$}&
      \multicolumn{1}{c}{$k_{(m,N)}\leta$}&
      \multicolumn{1}{c}{$\deltax_{(c)}/\leta$}&
      \multicolumn{1}{c}{$\deltax_{(c)}/\lc$}&
      \multicolumn{1}{c}{$\ReLeddy$}&
      \multicolumn{1}{c}{$\Relambda$}&
      \multicolumn{1}{c}{$\Lxyz/\leta$}&
      \multicolumn{1}{c}{$\Leddy/\leta$}&
      \multicolumn{1}{c}{$\lambdaTaylor/\leta$}\\[2.ex]
      \DNS & $128$ & $\np{\kfEta}$ & $\np{\kmaxEta}$ & $\np{\dxEta}$ & $\np{\dxlc}$ &
      $\np{\NHSReLeddy}$ & $\np{\NHSRelambda}$ &
      $\np{\Loleta}$ & $\np{\Leddyoeta}$ &
      $\np{\lambdaoleta}$ \\[2.0ex] 
      $\nc=\,4$ & $32$ & $\np{\kfEta}$ & \np{\kNyquistetaOne} & $\np{\dxcEtaOne}$ & $\np{\dxlcOne}$ &
      & &
      & &
      \\[1.ex]
      $\nc=\,8$ & $16$ & $\np{\kfEta}$ & \np{\kNyquistetaTwo} & $\np{\dxcEtaTwo}$ & $\np{\dxlcTwo}$ &
      & &
      & &
      \\[1.ex]
      $\nc=16$ & $8$ & $\np{\kfEta}$ & \np{\kNyquistetaThree} & $\np{\dxcEtaThree}$ & $\np{\dxlcThree}$ &
      & &
      & &
      \\[1.ex]
    \end{tabular}
    \caption{
      Physical and numerical parameters of the DNS database used this study, together with information on the low-resolution fields (coarsening factor $\equiv M$).
      $\Ni$ denotes the number of grid nodes per spatial direction and
      $\kmax$ is the largest resolved wavenumber, which we scale here with the Kolmogorov length $\leta$.
      For the coarsened fields, we report the Nyquist wavenumber $\kNyquist$ scaled by $\leta$.
      $\deltax$ (DNS run) and $\deltax_{c}$ (coarsened fields) are the grid spacings in the DNS ground truth and the coarsened fields, respectively, while $\lc$ is the critical length scale for turbulence synchronisation.
      For the full DNS, we also provide values for the large-eddy ($\ReLeddy$) and the Taylor Reynolds numbers ($\Relambda$).
      The last three columns summarise some key length scale ratios.
    }
    \label{tab:param_phys_num}
  \end{center}
  \vspace{-2ex}
  \hrulefill
\end{table}

%% file: sections/numerics.tex
We describe here the differentiable DNS solver used in this study, along with the candidate state-estimation procedures which are the focus of this paper. These include the classic \DA{} algorithm and the machine learning `super-resolution' approach inspired by variational data assimilation.

\subsection{DNS code}
The pseudo-spectral DNS solver used in this study is an adaptation of the two-dimensional, spectral version of the JAX-CFD solver \citep{Dresdner_al_2022}.
We implement the three-dimensional problem using the velocity-vorticity formulation proposed by \citet{Kim_Moin_Moser_1987}, maintaining the underlying GPU efficiency of the JAX-CFD timestepping routines and associated data structures \citep{Kochkov_al_2021,Dresdner_al_2022}, which was built around the JAX library \citep{Jax_github}.
The differentiability of the two-dimensional version of the code has previously been used to find unstable periodic orbits \citep{page2022recurrent,Page2025rev} and for the design of mixing strategies in complex fluids \citep{Alhashim2025}.

The governing equations in the velocity/normal vorticity formulation of \citet{Kim_Moin_Moser_1987} are
\begin{equation}
  \partial_{t}\, \qvec
  =\; \hvec + \dfrac{1}{\Regen} \Delta\, \qvec \hspace{10ex}
\end{equation}
where $\qvec\coloneq(\Delta v,\omy)^T$ and $\hvec\coloneq(h_v,h_{\omega})^T$ with
\begin{alignat}{2}
  h_v
  &=\; - \partial_{y}
        \left(
        \partial_{x}{H_1} + \partial_{z}{H_3}
        \right) +
        \left(
        \partial_{x}^{2} + \partial_{z}^{2}
        \right) H_2
        \label{eq:theory_KMM_hv},\\[1.0ex]
  h_{\omega}
  &=\;  \partial_{z}{H_1} - \partial_{x}{H_3}.
        \label{eq:theory_KMM_homy}
\end{alignat}
In equations~\eqref{eq:theory_KMM_hv} and \eqref{eq:theory_KMM_homy}, the term
$\Hvec = (H_1,H_2,H_3)^T \coloneq -(\uvec\boldsymbol{\cdot}\nablavec)\uvec + \fvecext$
combines the nonlinear convective term and the external body force $\fvecext$.
Both $\uvec$ and $\omvec$ are expanded as truncated Fourier series in all three spatial dimensions and are de-aliased according to the $2/3$-rule.
The nonlinear ($\Hvec$) terms are advanced using a $4$th-order Runge Kutta method, while the diffusion is treated with a semi-implicit Crank-Nicholson scheme.

%
\input{sections/images/omy_xyplanes_snaps_DNSlongtime.tex}
%
In all simulations, the velocity field is discretized by $\Nx\times\Ny\times\Nz=128^3$ collocation points, with a maximum resolved wavenumber $\kmax\leta=\np{\kmaxEta}$.
Three different coarse-graining factors $\nc\in\{4,8,16\}$ between the fine \DNS{} grid and the coarser observation grids are considered for the state estimation problem -- while the lower two values fall below the synchronisation limit $\lc=5\pi\leta\approx\np{\lcdx}\deltax$ \citep{Lalescu_Meneveau_Eyink_2013},
the most severe coarsening leads to a grid width about $40\%$ larger than $\lc$ (see also \autoref{tab:param_phys_num}).
In \autoref{fig:omy_xyplanes_snaps_DNSlongtime}, the coarsened grids are visualised in front of representative high-resolution snapshots of the vorticity, together with the Taylor microscale $\lambdaTaylor$ and the synchronisation limit $\lc$. At $\nc=4$, the mesh is sufficiently fine to resolve
the vortical structures visible in the figure, while $\nc=16$ is visibly much wider than the individual flow structures.

\subsection{Data assimilation}
Considering the Navier-Stokes equations
as a perfect forecasting model, a classical \DA{} approach seeks an initial flow state $\vvec_0\in\physspace$ in the physical `target' space $\physspace$ that minimises a loss function:
\begin{equation}
  \LossfctDA(\vvec_0) = \dfrac{1}{\NT} \displaystyle\sum_{k=0}^{\NT-1}
                        \norm{\yvec_{t_k} - \observop \circ \tfm{t_k}(\vvec_0)}^2,
  \label{eq:LossfctDAgen}
\end{equation}
where $\tfm{t}(\uvec)$ is the time forward map associated with the governing equations (which in our formulation includes back-and-forth conversion from primitive variables to the velocity/vorticity form), the variable $\yvec_{t_k}\in\obsspace$ corresponds to one of $\NT$ observations ($\obsspace$ is an abstract space of observations), and $\observop:\physspace\to\obsspace$ maps the velocity field into observations.
Assimilation is performed over a time window $t_k\in[0,\TunrollDA]$, and the objective of minimising \eqref{eq:LossfctDAgen} corresponds to finding an initial condition $\vvec_0$ that best reproduces the time-series of measurements as it is unrolled in time.

In the current study, the coarse-grained velocity fields from the DNS serve as the observations, and the observation operator ($\observop$ above) is a simple downsampling operation which we denote $\coarseop$.
The physical space $\physspace = \realn^{\Nx \times \Ny \times \Nz \times 3}$ includes our high-resolution fields, while the observation space $\obsspace = \realn^{\Nx/\nc \times \Ny/\nc \times \Nz/\nc \times 3}$ contains the corresponding coarse-grained fields.
The optimisation problem~\eqref{eq:LossfctDAgen} therefore becomes
\begin{equation}
  \LossfctDA(\vvec_0) = \dfrac{1}{\NT} \displaystyle\sum_{k=0}^{\NT-1}
                        \norm{\coarseop \circ \tfm{t_k}(\uvec_0) -
                              \coarseop \circ \tfm{t_k}(\vvec_0)}^2,
  \label{eq:LossfctDA}
\end{equation}
given a time series of coarse-grained ground truth snapshots $\{\coarseop \circ \tfm{t_k}(\uvec_0)\}_{k=0}^{\NT-1}$ as observations.
We compute gradients of the loss (via automatic differentiation) with respect to the initial condition, $\nablavec_{\vvec_0} \LossfctDA$, and supply them to an Adam optimiser \citep{Kingma_Ba_2015} with initial learning rate $\lrate=0.1$. In all computations, we take $200$ gradient update steps and select the best-performing $\vvec_0$.
The choice of the initial guess for the optimisation procedure will be discussed in the next section.
%

\subsection{Super-resolution}
Standard super-resolution approaches train a neural network to generate high-resolution images (here velocity fields from a high-fidelity DNS run) from low-resolution datasets. The model thus learns an interpolation operator $\nnetop:\obsspace\to\physspace$ (i.e. the convolutional neural network) as an approximate inverse to the coarse-graining operator $\coarseop$.
The variable $\nnetweigts$ indicates the network weights that minimise a loss of the form \citep{Fukami_Fukagata_Taira_2019,Fukami_Fukagata_Taira_2021}:
\begin{equation}
  \LossfctSR = \dfrac{1}{\NS} \displaystyle\sum_{j=1}^{\NS}
                 \norm{\uvecj -
                       \nnetop \circ \coarseop(\uvecj)}^2.
  \label{eq:LossfctSR}
\end{equation}
Training is performed on $\NS$ individual snapshots (here velocity fields $\uvecj$).
As discussed in \S\ref{sec:intro}, training a neural network with the loss~\eqref{eq:LossfctSR} can lead to `unphysical' velocity fields, e.g. not divergence free.
Therefore, additional terms or modifications to the loss are typically required to ensure that the reconstructed fields fulfil at least some of the desired physical properties (see \citealp{Fukami_Fukagata_Taira_2023} and references therein).

In analogy to the \DA{} loss function~\eqref{eq:LossfctDA} above, the loss in our trajectory-based super-resolution technique (hereafter `\SRdyn{}') contains low-resolution trajectories that are obtained from unrolling the coarse-grained training dataset over a fixed unroll time $\Tunroll$ \citep{Page_2025}:
\begin{equation}
  \LossfctTC = \dfrac{1}{\NS\NT} \displaystyle\sum_{j=1}^{\NS} \displaystyle\sum_{k=0}^{\NT-1}
               \norm{\coarseop \circ \tfm{t_k}(\uvecj) -
                     \coarseop \circ \tfm{t_k} \circ \nnetop \circ \coarseop(\uvecj)}^2.
  \label{eq:LossfctTC}
\end{equation}
The loss~\eqref{eq:LossfctTC} measures the deviation between a time series of coarse-grained ground truth observations $\{\coarseop \circ \tfm{t_k}(\uvecj)\}_{k=0}^{\NT-1}$ and the coarse-grained trajectory obtained by time-marching the
high-resolution fields produced by the neural network and coarsening the output.
The deviation from the reference observations is evaluated at $\NT$ discrete times $t_k \in \{0, \deltatc, \dots, (\NT - 1)\deltatc\}$ along each trajectory in the interval $[0,\Tunroll]$, where $\deltatc = \nct\deltat$ denotes the time interval between two consecutive observation times and $\deltat$ is the simulation timestep in the DNS.
We choose the temporal coarsening factor to match the degree of spatial coarse-graining, $\nct=\nc$, to mimic the limited temporal availability of observational data, for instance, in an experiment.
The neural network includes a Leray projection $\Lerayop(\widetilde{\uvec}) \coloneq \widetilde{\uvec} - \nablavec\Delta^{-1}\nablavec\boldsymbol{\cdot}\widetilde{\uvec}$ as its final `layer' to guarantee its output to be divergence free.

\nprounddigits{0}
\FPeval\TtotPerTraj{round(100/\NHSteddy:0)}
\FPeval\dTPerTraj{round(2/\NHSteddy:0)}
\FPeval\Tunrollone{round(0.25/\NHSteddy:2)}
\FPeval\Tunrolltwo{round(0.50/\NHSteddy:2)}
\FPeval\Tunrollthree{round(1.50/\NHSteddy:2)}
All \SRdyn{} models were trained to minimise the loss~\eqref{eq:LossfctTC} using a database of $\Ntraj=75$ independent trajectories, each of length $\TtotPerTraj\teddy$ (equivalent to $100$ non-dimensional time units), with $50$ snapshots stored per trajectory ($3750$ total snapshots).
We remove $10\%$ of the training dataset for validation (to verify the model is not over-fitting).
The results presented in \S\ref{sec:results} are based on a separate `test' set of velocity field snapshots obtained in the same way.
Training is performed with an Adam optimiser \citep{Kingma_Ba_2015} with a learning rate $\lrate=10^{-4}$.
We train for between 25 and 50 epochs (complete cycles through the training data) with a batch size of $16$ individual trajectories (note the ability to run many independent DNS calculations simultaneously on a GPU).
We explored several unroll times in a range $\Tunroll/\teddy\in[\Tunrollone,\Tunrollthree]$, which indicated a rather weak impact of this parameter on the performance of the trained models.
Given this and the fact that the time-marching of the DNS is the most intensive aspect of the training loop, our results reported here are mostly for networks trained with an unroll time of $\Tunroll=\Tunrollone\teddy$ (equivalent to $0.25$ non-dimensional time units).

%
\input{sections/images/sketch_network_architecture.tex}
%
The overall architecture of the neural networks depicted in \autoref{fig:sketch_network_architecture} is conceptually similar to the one applied by \JacobJFM{} in two-dimensional Kolmogorov flow.
The networks are purely convolutional and feature a residual network structure (`ResNet', \citealp{He_Zhang_Ren_Sun_2016}):
the ResNet `layer' learns a correction to the input, $\avec$, which is then added to the input $\avec\to\mathcal{F}(\avec)+\avec$ ($\mathcal{F}$ is the learned operation).
Each `Upsampling + residual block' operation (see figure \ref{fig:sketch_network_architecture}) doubles the size of the input in all three spatial directions.
Hence, for a coarse-graining factor $\nc=2^n$, we include $n$ individual residual layers to eventually reconstruct a field at the target resolution.
All convolutional layers are fully three-dimensional and feature $32$ filters with either GeLU \citep{Hendrycks_Gimpel_2016} or linear (no) activation functions.
The kernel size is fixed at $4\times 4\times 4$, while periodic padding is applied along all snapshot boundaries.
The Leray projection discussed above forms the last layer in the model and ensures that the obtained high-resolution field is solenoidal.
The entire training process is implemented using the Keras library with a JAX backend \citep{Chollet_2015} to
allow for straightforward inclusion of the JAX-CFD-based DNS solver and the ability to perform end-to-end differentiation in a single line of code.
Note that an extensive variation and optimisation of the model architecture is beyond the scope of this study. As such there will likely be some room for further improvement in the predictive capability of the model.

We found that pre-training the network for $50$ epochs
to reproduce an interpolated high-resolution field is highly beneficial to the final model performance:
\begin{equation}
  \LossfctInt = \dfrac{1}{\NS} \displaystyle\sum_{j=1}^{\NS}
                \norm{\interpop \circ \coarseop(\uvecj) -
                      \nnetop \circ \coarseop(\uvecj)}^2,
\end{equation}
where $\interpop$ represents tri-cubic spline interpolation.

%% file: sections/images/omy_xyplanes_snaps_DNSlongtime.tex
\begin{figure}
  \centering
  \includegraphics[width=\linewidth]
  {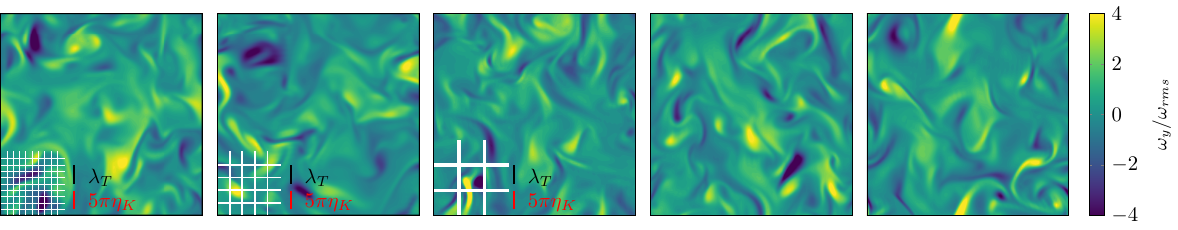}
  \caption{
    Sample slices of the vorticity field $\omy/\omrms$ at $z=0$.
    Black and red lines indicate
    the Taylor microscale $\lambdaTaylor$ and
    the critical length scale $\lc=5\pi\leta$, respectively.
    The white grid lines in the lower left corners of the first three frames visualise the coarsened grids
    for coarsening factors $\nc\in\{4,8,16\}$ relative to the DNS grid.
    }
\label{fig:omy_xyplanes_snaps_DNSlongtime}
\end{figure}

%% file: sections/images/sketch_network_architecture.tex
\begin{figure}
  \centering
  \includegraphics[width=0.95\linewidth]
  {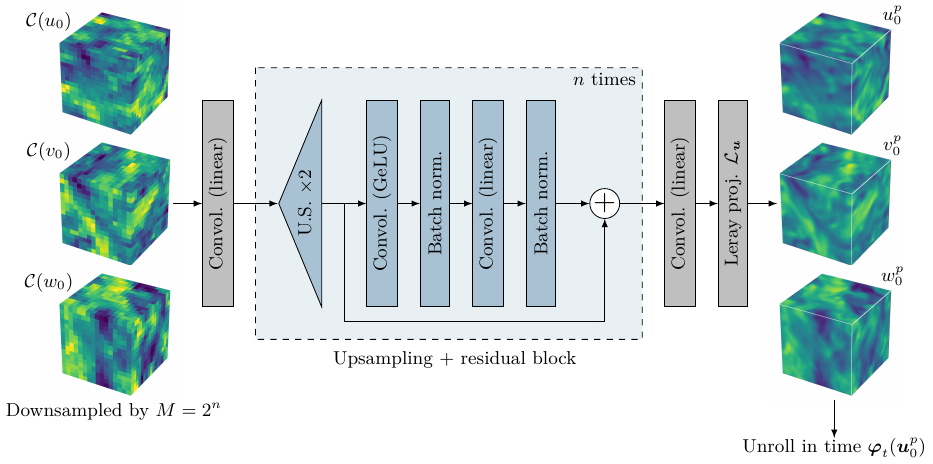}
  \caption{
    Neural network architecture used to train the \SRdyn{} model with `U.S.' indicating an upsampling layer.
    The field $\uvecpredzero$ that the network reconstructs from the coarse-grained DNS snapshot $\coarseop(\uveczero)$ is subsequently advanced in time to compute the loss function~\eqref{eq:LossfctTC}.
  }
\label{fig:sketch_network_architecture}
\end{figure}

%% file: sections/results.tex
\subsection{Super-resolution with a dynamic loss}\label{subsec:SRdyn}
We first analyse the predictive performance of the three-dimensional \SRdyn{} technique and compare it with both simple polynomial interpolation and standard \DA{}.
Coarse-grained fields $\coarseop(\uveczero)$ are given to the different state estimation techniques and their reconstructions $\uvecpredzero$ are compared to the ground truth state.
An initial guess for the \DA{} algorithm is generated by interpolating the coarse-grained field onto the high-resolution grid using tri-cubic interpolation, $\interpop\circ\coarseop(\uveczero)$.

\input{sections/images/errL2uvw_vs_time.tex}
%
\emph{State reconstruction and performance under time marching:}
To quantify the deviation between predicted and observed trajectories, we first compute the relative reconstruction error
\begin{equation}
  \erruvw(t) = \dfrac{ \norm{ \tfm{t}(\uveczero) - \tfm{t}({\uvecpredzero})}}
                  { \norm{\tfm{t}(\uveczero)} },
  \label{eq:errL2uvw}
\end{equation}
for the various state estimation schemes (4DVar, \SRdyn{} and simple polynomial interpolation).
The temporal evolution of (\ref{eq:errL2uvw}) is reported for the different state estimation schemes in \autoref{fig:errL2uvw_vs_time}.
The shaded regions in this figure correspond to $\pm$ one standard deviation around the mean for (i) the full test dataset in \SRdyn{} (grey) and (ii) 16 independent initial conditions for 4DVar (orange).
The solid lines are for a single initial condition -- which is representative of the test set as a whole -- that we examine in detail below.

Consistent with the findings of \JacobJFM{} in two-dimensional Kolmogorov flow, \SRdyn{} provides the most accurate prediction of the initial state and continues to outperform \DA{} for the initial $0.2$ to $0.4$ large-eddy times, depending on the degree of coarse-graining. With time, the trajectories predicted by classical \DA{} approach the ground truth trajectory and finally reach the lowest reconstruction error around the end of their respective assimilation windows.
At these later times, \DA{} strongly benefits from its optimisation which incorporates observations of the specific ground truth trajectory to which the neural network has no access: The neural network instead generates the most plausible high-resolution state from a single coarse-grained field based on its training experience with a large number of coarse-grained evolutions.
In this regard, it is remarkable that the time-marched \SRdyn{} state leads to a comparably good or even better reconstruction of the target trajectory than some of the \DA{} runs, at least for $\nc=4$ and $\nc=16$.
As expected, on a time scale clearly longer than $\TunrollDA{}$, the reconstruction error of all state estimation techniques rises rapidly and no reliable predictions are possible over these time horizons.
As expected, tri-cubic interpolation consistently leads to the highest reconstruction errors; it is comparable in performance to 4DVar only at the harshest coarse-graining and then only at $t=0$.

\input{sections/images/v_xyplanes_snaps_T0.tex}
%
\emph{Instantaneous velocity fields:}
A visual comparison of reconstructed velocity fields for the example initial condition of figure \ref{fig:errL2uvw_vs_time} (solid lines in the figure) is reported in \autoref{fig:v_xyplanes_snaps_T0} for each of the three coarsening factors $\nc\in\{4,8,16\}$.
All three techniques appear to do well, at least qualitatively, at the lowest coarse-graining factor $M=4$, which is unsurprising given the favourable comparison to the synchronisation scale, $\Delta x_{c} \approx 0.35 \lc$.
When the coarse-graining factor is raised to $\nc=8$ ($\Delta x_{c}\approx0.69\lc$) and $\nc=16$ ($\Delta x_{c}\approx 1.38\lc$), resolution of typical vortical structures is lost (see figure \ref{fig:omy_xyplanes_snaps_DNSlongtime}).
As a result, tri-cubic interpolation of the under-resolved velocity fields fails to correctly identify the shape and physical location of the intense velocity fluctuations -- see the red boxes in figure \ref{fig:v_xyplanes_snaps_T0} highlighting particular structures.
The learned interpolation operator $\nnetop$ reduces the exaggerated smoothing effect of simple interpolation and retains -- in contrast to both 4DVar and polynomial interpolation -- the ability to correctly reconstruct the location of the individual intense velocity patterns (examples highlighted in red boxes), even beyond the synchronisation limit $\lc$.
The 4DVar reconstructions are increasingly dominated by high-wavenumber oscillations and feature clearly overestimated peak values at the larger coarse-graining values.

\input{sections/images/uvw_omx_eps_pdf_var_time.tex}
%
Corresponding probability density functions (PDF) of the velocity, vorticity and dissipation rate are shown in \autoref{fig:uvw_omx_eps_pdf_var_time} (upper rows in each panel (a,b,c) correspond to the snapshots of figure \ref{fig:v_xyplanes_snaps_T0}, lower rows are computed at $t=\teddy$).
For the mild coarse-graining $\nc=4$, all three techniques reproduce the PDF of the velocity fluctuations $\uprime$ at $t=0$ almost perfectly, with slight deviations in the tails of the \DA{}-based reconstruction.
On the other hand, \DA{} overestimates the tails of the PDFs for both vorticity fluctuations $\omprime$ and the local dissipation rate $\eps$, while underestimating the more probable events. Classical polynomial interpolation and \SRdyn{} provide a much better reconstruction of the ground truth PDFs, with only the most extreme values being under-represented in the reconstructed field.

Consistent with the observations in \autoref{fig:v_xyplanes_snaps_T0}, \DA{}'s tendency to overestimate the PDF of the velocity fluctuations increases with further coarse-graining, while \SRdyn{} (and also the simple polynomial interpolation) tends to smooth out the extreme values in the field more strongly.
The same trend is seen in the PDFs of $\omprime$ and $\eps$, but the deviations from the target PDF are more pronounced.
The distributions of the \DA{}-reconstructed fields are much wider than that of the ground truth state -- the assimilated fields over-estimate the occurrence of extreme gradients.
On the other hand, the low-amplitude values are over-represented in the fields generated via simple polynomial interpolation.
The super-resolution approach \SRdyn{} consistently provides the most faithful reconstruction of the PDFs, although its deviations are still significant for $\nc=16$.
Given the poor performance of simple tri-cubic interpolation across the various metrics analysed so far, we do not consider this method further beyond this point.

Under time advancement, all PDFs evolve towards the ground truth as the corresponding trajectories rapidly collapse onto the turbulent attractor and the initial unphysical oscillations of \DA{} decay.
After one large-eddy time (lower rows of each panel in \autoref{fig:uvw_omx_eps_pdf_var_time}), all PDFs essentially collapse onto the ground truth for $\nc=4$ and $\nc=8$. For $\nc=16$, deviations especially in the PDF tails remain visible, though they are much weaker than for the initial reconstructed fields $\uvecpredzero$.

\input{sections/images/crosscorr_DNS_SRdyn_4DVar_var_time.tex}
%
%
In order to quantify how well the different scales of the velocity field are reconstructed by the different state estimation techniques, we introduce the normalised co-spectrum of $\uvec$ and $\uvecpred$ \citep{Li_Zhang_Dong_Abdullah_2020}
\begin{equation}
  \Crosscorr(k,t) = \dfrac{\displaystyle\int_{{\abs{\kxyz}=k}} d \kvec \ \uvecfour(\kvec,t) \uvecpredfour(\kvec,t)^{*}}
                          { 2\sqrt{\Ekspec_{\uvec}(k,t) \Ekspec_{\uvecpred}(k,t)} },
\end{equation}
where $\Ekspec_{\uvec}$ and $\Ekspec_{\uvecpred}$ are the energy spectra for a ground truth state and its reconstruction, respectively.
Hence, $\Crosscorr$ measures the correlation between a ground truth Fourier mode $\uvecfour$ and its reconstructed counterpart $\uvecpredfour$ for each wavenumber $k$, with a value close to unity (zero) indicating a very good (poor) reconstruction of the respective Fourier mode.

\FPeval\kNyquistkcOne{\kNyquistetaOne/0.2}
\FPeval\kNyquistkcTwo{\kNyquistetaTwo/0.2}
\FPeval\kNyquistkcThree{\kNyquistetaThree/0.2}
\nprounddigits{1}
In \autoref{fig:crosscorr_DNS_SRdyn_4DVar_var_time}, $\Crosscorr$ is evaluated for the different coarse-graining factors at times $t/\teddy\in\{0,0.5,1\}$ along the trajectories.
The Nyquist cutoff wavenumbers of the coarse-graining operator are larger than the synchronisation value at $M=4$ and $M=8$ ($\kNyquist \approx \np{\kNyquistkcOne}\kc$ and $\kNyquist \approx \np{\kNyquistkcTwo}\kc$ respectively) but smaller at $M=16$ ($\kNyquist \approx \np{\kNyquistkcThree}\kc$).
In the \SRdyn{}-generated reconstruction, all scales up to the Nyquist wavenumber are more or less perfectly reproduced for the different coarsening factors, followed by a range of partially recovered modes $k>\kNyquist$ whose number reduces with increasing $\nc$.
\DA{}, on the other hand, has difficulty reconstructing the large scales of the initial velocity field, with markedly reduced values of $\Crosscorr$ appearing before the Nyquist cutoff. For the strongest coarse graining $\nc=16$, standard \DA{} even struggles to accurately reproduce the largest scales into which energy is injected by the external forcing.

\nprounddigits{1}
As the reconstructed fields are unrolled for $\nc=4$ (\autoref{fig:crosscorr_DNS_SRdyn_4DVar_var_time}\textit{a}), both techniques provide accurate reproductions of all dynamically relevant scales, with \SRdyn{} even outperforming some of the \DA{} runs up to $t=\teddy$.
For $\nc=8$ (\autoref{fig:crosscorr_DNS_SRdyn_4DVar_var_time}\textit{b}), \SRdyn{} is still able to faithfully reproduce the large-scale content of the velocity field over a large-eddy time, but the high-wavenumber modes deviate from those in the ground truth trajectory as time increases. This is not unexpected since the smallest scales are known to decorrelate much faster than their large-scale counterparts \citep{Boffetta_Musacchio_2017}.
The \DA{} runs, on the other hand, are optimised to match the coarse-grained observations as closely as possible and provide better approximations for a wider range of scales when $t\geq0.5\teddy$.
Finally, with $\kNyquist$ falling well below $\kc$ at $M=16$
(\autoref{fig:crosscorr_DNS_SRdyn_4DVar_var_time}\textit{c}), even \DA{} is not capable of reproducing much more than the largest flow scales in a satisfactory way.
Notably, the high-resolution fields obtained by \SRdyn{} again outperform all of their \DA{}-generated counterparts in the early stages of the time evolution and continue performing equally well as the \DA{} run with the shortest assimilation window, i.e. $\TunrollDA\approx\np{\TunrollNormA}\teddy$.
\nprounddigits{2}

Our observations on the performance of \DA{}, in particular the decreasing quality with increasing coarse-graining, is in line with what is known about assimilation in the context of turbulence synchronisation (see discussion in \S\ref{sec:intro}) and is in qualitative agreement with the results of \citet{Li_Zhang_Dong_Abdullah_2020} who performed a similar \DA{}-based study in a triply-periodic flow with monochromatic `Kolmogorov' forcing (which drives a large-scale mean flow).
A one-to-one comparison with the latter study is not straightforward due to the change in forcing profile and the fact that their \DA{} algorithm operates in spectral space, where the leading Fourier modes of the ground truth at wavenumbers $0\leq k\leq k_O$ act as observations for the optimisation.
For comparison, we performed some \DA{} runs for the Kolmogorov forcing used by \citet{Li_Zhang_Dong_Abdullah_2020} at identical parameters -- the key difference is that the optimisation is performed in physical space and exclusively with coarse-grained observational data.
These calculations are included in \autoref{sec:appendix3DKolmogorov}.
In line with previous observations in other data assimilation techniques \citep{DiLeoni_Mazzino_Biferale_2020},
the comparison reveals that, for a Nyquist cutoff $\kNyquist$ matching the spectral cutoff wavenumber $k_O$, \DA{} based on coarse-grained physical observations leads to weaker reconstructions than the purely spectral procedure of \citet{Li_Zhang_Dong_Abdullah_2020}.

\input{sections/images/omy_xyplanes_snaps_DNS_SRdyn_4DVar_nc8_vs_time.tex}
%
\emph{Reproduction of vortical structures:}
Snapshots of the instantaneous vorticity field derived from the reconstructed state, $\omvecpred(\xvec,t)\coloneq\nablavec\times\tfm{t}(\uvecpredzero)$, are shown in \autoref{fig:omy_xyplanes_snaps_DNS_SRdyn_4DVar_nc8_vs_time} for the intermediate coarse-graining level $\nc=8$ alongside the corresponding ground truth states.
Given that both neural network training and data assimilation have been performed based on velocity field snapshots, reconstructing gradient-based fields like $\omvecpred$ is an additional challenge.
The \SRdyn{}-based reconstruction of the initial vorticity field $\omvecpredzero\coloneq\omvecpred(\xvec,0)$ reveals a qualitatively similar pattern of small-scale structures to the target DNS snapshot, albeit contaminated with some background noise.
The vorticity field reconstructed by standard \DA{}, on the other hand, deviates significantly from the target state in terms of both the spatial organisation and amplitude of the intense vorticity regions, consistent with the overestimated tails of the corresponding PDF in \autoref{fig:uvw_omx_eps_pdf_var_time}.

When advanced in time, the strong initial deviations between the \DA{}-reconstructed field and the DNS reference state decay rapidly, as does the weaker background noise in the \SRdyn{}-based reconstruction.
At $t=0.25\teddy$, the deviations of the \SRdyn{}-based reconstruction and the ground truth are minimal (see the highlighted regions in \autoref{fig:omy_xyplanes_snaps_DNS_SRdyn_4DVar_nc8_vs_time}).
While \DA{} also provides a reasonable reconstruction of the vorticity field, the deviations from the ground truth are more clearly discernible.
This situation changes at $t=0.5\teddy$, where \DA{} now provides a slightly more accurate reconstruction of the ground truth, though the \SRdyn{}-based reconstructions still perform well considering that they have not seen observational data of the specific trajectory.
For $t>0.5\teddy$, deviations between the \SRdyn{}-generated reconstructions and the ground truth snapshots grow stronger -- at $t=\teddy$, only the large-scale organisation of the vorticity fields are still comparable.
A similar deviation from the ground truth trajectory occurs for the \DA{}-based reconstruction once the trajectory leaves the assimilation window (see snapshots at $t=2T_e$ in \autoref{fig:omy_xyplanes_snaps_DNS_SRdyn_4DVar_nc8_vs_time}).

\input{sections/images/qhunt_3diso_DNS_SRdyn_4DVar_nc8_vs_time.tex}
%
A complementary view is provided in \autoref{fig:qhunt_3diso_DNS_SRdyn_4DVar_nc8_vs_time}, where three-dimensional organisation of vortical structures visualised using isosurfaces of the second invariant of the velocity gradient tensor $Q$ \citep{Hunt_al_1988} are reported.
In agreement with our observations of vorticity slices in \autoref{fig:omy_xyplanes_snaps_DNS_SRdyn_4DVar_nc8_vs_time}, \SRdyn{} provides a reasonable reconstruction of the three-dimensional vortical structures in the initial field, though the field is less smooth than the ground truth due to the presence of low-amplitude background noise. For \DA{}, the range of values attained by $Q$ is seen to be greatly overestimated so that no individual features are visible with isosurfaces at $Q=\omrms^2$.
The initial low-amplitude reconstruction errors in the \SRdyn{}-generated state rapidly diffuse under time advancement and for $t=0.25\teddy$ and $t=0.5\teddy$, most vortices in the DNS target field are well captured by the neural network prediction (see the regions highlighted with red boxes in \autoref{fig:omy_xyplanes_snaps_DNS_SRdyn_4DVar_nc8_vs_time}).
For longer unroll times, the overall spatial organisation of the vortices remains comparable to the ground truth DNS, but the small-scale features in the \SRdyn{}-based reconstruction start to deviate.
The high errors of the \DA{} reconstruction also dissipate rapidly as the initial field is unrolled in time, and by $t=0.25\teddy$ the $Q$ isosurfaces largely match those in the reference trajectory.
However, it takes until $t=0.5\teddy$ for the \DA{}-based prediction to outperform \SRdyn{}. Towards the end of the assimilation window at $\TunrollDA\approx 1.3\teddy$, \DA{} reaches its best predictive performance (compare the minimum of $\erruvw$ in \autoref{fig:errL2uvw_vs_time}) and deviations from the ground truth are hardly visible.

\subsection{\DASR{} -- a hybrid approach}\label{subsec:DASR}
The results above show that the \SRdyn{} approach -- which does not rely on the availability of a library of high-resolution data -- can outperform 4DVar for state estimation in three-dimensional turbulence at $t=0$ (as measured by an $L_2$ norm to the ground truth), in line with the results reported in \JacobJFM{} for 2D Kolmogorov flow.

Similar to the 2D work, this performance does not continue to hold under time marching:
After an interval where both methods provide comparably good reconstructions of the target trajectory, standard \DA{} eventually wins thanks to its access to observations of the dynamics along the specific trajectory.
One well-known challenge in the non-linear optimisation framework of \DA{} is the non-uniqueness of the target high-resolution field -- several trajectories starting from very different states $\uvecpredzero$ may reconstruct the coarse observations at late times with similar accuracy \citep{Zaki_2025}.

One approach to improve the poor reconstruction of early-time states in \DA{} is to emphasise reconstruction of early observations by adding time-dependent weights to the loss function \citep{Wang_Wang_Zaki_2019}.
Here, we instead aim at improving the quality of the prediction by initialising the \DA{} optimisation algorithm with a more accurate, neural network-generated reconstruction of $\uveczero$ as the initial guess, all without requiring any knowledge of high-resolution fields.
Our approach (hereafter termed `\DASR{}') shares some similarities with the method of \citet{Frerix_al_2021}, who used a neural network trained on high-resolution data in two-dimensional Kolmogorov flows to initialise \DA{}.

%
\input{sections/images/v_xyplanes_snaps_T0_DNS_SRdyn_4DVar_4DVarSR.tex}
%
%
The positive influence of the \SRdyn{}-based initialization on the \DA{} reconstruction of $\uveczero$ is demonstrated in \autoref{fig:v_xyplanes_snaps_T0_DNS_SRdyn_4DVar_4DVarSR},
where the \DASR{} field $\uvecpredzero$ is shown alongside the results obtained with \SRdyn{} and \DA{} at the more challenging coarse-graining levels $\nc=8$ and $\nc=16$.
The \DASR{}-generated prediction features much less contamination with erroneous small-scale features than in standard \DA{}, while the predicted values fall in a similar range as those in the ground truth.
The hybrid \DASR{} approach is also able to correctly locate the high-speed regions marked by red boxes for $\nc=16$, in contrast to standard \DA{} which predicts a state in which the most intense velocity patches are shifted with respect to the DNS ground truth.
At the less-demanding value of $M=8$, a comparison of the \DASR{}- with the \SRdyn{}-based reconstruction reveals only minor deviations.
This is not unexpected since the \SRdyn{} is used to initialize the \DASR{} procedure.
However, at the strongest coarse-graining $\nc=16$, the \DASR{} high-resolution snapshot $\uvecpredzero$ differs more strongly from \SRdyn{}.
Some of the erroneous small-scale features characteristic of \DA{} have emerged during the optimisation, but at a much weaker amplitude so that they do not dominate the reconstructed field.

\input{sections/images/uvw_omx_eps_pdf_DNS_4DVarSR_nc8_nc16_var_time.tex}
%
\autoref{fig:uvw_omx_eps_pdf_DNS_4DVarSR_nc8_nc16_var_time} presents standardised PDFs of velocity and vorticity fluctuations as well as fluctuations of the local dissipation rate $\eps$, for $M=8$ shown in subfigure (a) and $M=16$ in subfigure (b). In both subfigures, the upper rows correspond to data at $t=0$ and the bottom rows to data at $t=T_e$.
In support of the previous observations, the \DASR{} prediction of the PDFs at $t=0$ represents a significant improvement on those obtained with `standard' \DA{} and \SRdyn{} (latter not shown here, see \autoref{fig:uvw_omx_eps_pdf_var_time}).
The results are particularly impressive for the two gradient-based quantities $\omprime$ and $\eps$, for which standard \DA{} greatly over-predicts the tails of the distribution and \SRdyn{} misses the most extreme values.
The \DASR{} PDFs almost perfectly reproduce the DNS ground truth data at the intermediate coarse graining level $\nc=8$, and clearly reduce the over-prediction of the PDF tails compared to the standard \DA{} at the more demanding $M=16$.
Unrolling the reconstructed fields over one large-eddy time
leads to an almost perfect match between \DASR{} and the PDFs of the ground truth even at $\nc=16$, while standard \DA{} still underestimates the probability of intense vorticity and dissipation events, as can be seen from the data shown in rows two and four of \autoref{fig:uvw_omx_eps_pdf_DNS_4DVarSR_nc8_nc16_var_time}.

\input{sections/images/errL2uvw_4DVarSR_varTunroll_vs_time.tex}
%
In \autoref{fig:errL2uvw_4DVarSR_varTunroll_vs_time}, the evolution of the pointwise relative reconstruction error $\erruvw$ is compared for \DA{} and \DASR{} for an example trajectory.
Initializing the data assimilation procedure with the neural network prediction reduces the initial error $\erruvw(0)$ by more than factor two compared to standard \DA{}.
Beyond $t=0$, the initialization with the \SRdyn{}-generated field has a lasting influence on the quality of the prediction over the entire observation interval, allowing \DASR{} to outperform the classical \DA{} predictions for all coarse-graining levels and assimilation window lengths. Moreover, for both $\nc=4$ and $\nc=8$, the error growth rate after passing the overall minimum is noticeably reduced compared to the \DA{} runs -- reliable predictions are possible over longer time intervals.

\input{sections/images/crosscorr_DNS_4DVar_4DVarSR_var_time.tex}
%
The Fourier correlation coefficient $\Crosscorr(k,t)$ reported in \autoref{fig:crosscorr_DNS_4DVar_4DVarSR_var_time} shows that the significantly lower reconstruction errors in \DASR{} are associated with a more accurate reproduction of a wider range of wavenumbers in the initial field $\uveczero$ than both \SRdyn{} and `standard' \DA{}.
The \DASR{} method has further improved the \SRdyn{} prediction of $\uveczero$, providing more faithful reconstructions of the modes with $k>\kNyquist$.
As the \DASR{} fields are unrolled in time, the resulting trajectory follows the ground truth more robustly than \DA{}: even at $t=\teddy$, we observe mode-wise correlations of more than $75\%$ for all resolved wavenumbers.
Later at $t=2\teddy$, which is well beyond the end of the assimilation window $\TunrollDA\approx\TunrollNormC\teddy$, the \DASR{} reconstruction of the low-wavenumber content is still fairly good at both coarse-graining levels.
For $\nc=8$, this holds even for most of the high-wavenumber modes, and snapshots of both the velocity and vorticity fields are still almost indistinguishable from their ground truth counterparts (not shown).

\input{sections/images/omy_xyplanes_snaps_DNS_4DVar_4DVarSR_nc16_vs_time.tex}
%
For the strongest coarse-graining, \autoref{fig:omy_xyplanes_snaps_DNS_4DVar_4DVarSR_nc16_vs_time} shows selected snapshots of the \DA{}- and \DASR{}-generated vorticity $\omvecpred$, alongside the corresponding states of the ground truth DNS.
The initial field reconstructed by means of standard \DA{} shares very little similarity with the ground truth, with the range of reconstructed values far exceeding that of the target field.
The range of values attained in the \DASR{}-reconstructed field is much closer to that of the DNS reference state, but still features a pronounced unphysical high-wavenumber noise, and only the large-scale pattern roughly matches the overall flow organisation in the target field.
For standard \DA{}, the vorticity amplitudes decay quickly with time and reach a range comparable to the ground truth state by $t=0.25\teddy$, though the spatial distribution of $\omypred$ remains fairly different from the corresponding ground truth state until $t=\teddy$.
Beyond the end of the assimilation window, the \DA{}-based reconstruction deviates strongly from the target field as the assimilated trajectory diverges from its ground truth counterpart.
For \DASR{}, on the other hand, the initial high-wavenumber noise has diffused away by $t=0.25\teddy$ and, from $t=0.5\teddy$ on, the vorticity field is visually almost indistinguishable from the ground truth up to a large-eddy time.

\input{sections/images/omy_xyplanes_errLocal_DNS_4DVar_4DVarSR_nc16_vs_time.tex}
%
\autoref{fig:omy_xyplanes_errLocal_DNS_4DVar_4DVarSR_nc16_vs_time} shows
the spatial distribution of the local reconstruction error $\omypred(\xvec,t)-\omy(\xvec,t)$ corresponding to the snapshots in
\autoref{fig:omy_xyplanes_snaps_DNS_4DVar_4DVarSR_nc16_vs_time}.
While standard \DA{} features errors across the entire domain for all snapshots (except for reasonable agreement at $t=\teddy$), the overall error amplitude for \DASR{} is much lower.
Deviations from the ground truth occur primarily in regions of intense vorticity.
A comparison with the `full' ground truth snapshots $\omy$ indicates that \DASR{} reproduces the spatial organisation of the vorticity field as a whole and that the error seems to originate in a mild under- or over-prediction of the amplitude in the patches of intense vorticity.
Beyond the end of the assimilation window (see snapshots at $t=2\teddy$), we find that the reconstruction error for \DASR{} does not grow uniformly in space.
Instead, individual flow features such as the highlighted intense vorticity patches are no longer captured in the reconstruction; for other flow features, the main error contribution seems to be an incorrectly predicted amplitude.

\input{sections/images/ek3d_4DVarSR_varTunroll_var_time.tex}
%
\autoref{fig:ek3d_4DVarSR_varTunroll_var_time} presents a comparison of \SRdyn{}, \DA{} and \DASR{} with respect to
their ability to faithfully reproduce the turbulent energy spectra of the target snapshots, with results pertaining to $\nc = 8$ and $\nc = 16$ shown in subfigures (a) and (b), respectively.
Results for $t=0$ are shown in the leftmost panels of both subfigures. As can be seen from the data shown in there, \SRdyn{} under-predicts the energy for $\kNyquist < k < 1/\leta$, while \DA{} over-predicts the energy levels for these wavenumbers.
The \DASR{}-based spectra fall in between and provide the best reconstruction of the ground truth state, mildly under-predicting the energy contributions at wavenumbers just above $\kNyquist$ and slightly over-predicting in the dissipative range. The high-$k$ over-estimation in \DASR{} and \SRdyn{} remains well below that of \DA{}.
This is consistent with our previous observations that \DASR{} significantly reduces the intense high-frequency noise in the initial field that was seen to characterise standard \DA{} at more intense coarse-graining.
The middle and right panels of \autoref{fig:ek3d_4DVarSR_varTunroll_var_time} correspond to measurements later in time, at $t=\teddy$ and $t = 1.5\teddy$, respectively. As can be seen from the data shown in these panels, the initial deviations in the mid- to high-wavenumber modes present in all reconstructions at $t=0$ quickly disappear as the fields are unrolled in time.
While the reconstructed spectra coincide with that of the ground truth at later times for $\nc=8$, both \SRdyn{} and \DA{} are seen to somewhat deviate from the target spectrum for $\nc=16$. \DASR{} again shows a better predictive performance and reproduces the ground truth almost perfectly, even for the most severe coarsening.

\input{sections/images/Pk3d_4DVarSR_varTunroll_var_time.tex}
%
\autoref{fig:Pk3d_4DVarSR_varTunroll_var_time} shows the corresponding instantaneous energy flux across wavenumber $k$, $\Pkspec(k,t)$, at three instances in time $t/\teddy\in\{0,1,1.5\}$ for $\nc=8$ (subfigure (a)) and $\nc=16$ (subfigure (b)).
As can be seen from the leftmost panel of subfigure (a), for the initial instant in time at the moderate coarse-graining level $\nc=8$, both standard \DA{} with $\TunrollDA/\teddy\in\{0.9,1.3\}$ and \SRdyn{}
reproduce the general shape of the flux for $k\leqslant \kNyquist$.
In this context, standard \DA{} with $\TunrollDA/\teddy = 0.9$ and $\TunrollDA/\teddy = 1.3$ is seen to perform well even at higher wavenumbers. Nonetheless, in comparison with the ground truth, the inter-scale energy flux is under-predicted by all methods in different wavenumber ranges (all wavenumbers in the case of standard \DA{} at $\TunrollDA\approx0.4\teddy$).
The \DASR{}-based reconstruction of $\Pkspec$ coincides very well with the DNS ground truth up to the Nyquist cutoff wavenumber $\kNyquist$, while predicting too low energy transfer rates towards higher wavenumbers.

The leftmost panel of \autoref{fig:Pk3d_4DVarSR_varTunroll_var_time}(b) presents results at $t=0$ for $\nc=16$. As can be seen from the data shown in the figure, \DASR{} (shown in red) is the only technique that is able to roughly reproduce the characteristic shape of the ground-truth flux, coinciding with the reference DNS for all $k < \kNyquist$.
The \DA{} and \SRdyn{} results, on the other hand, differ significantly in shape and amplitude from the ground truth.
\SRdyn{} strongly under-predicts the energy transfer to higher wavenumbers over the entire wavenumber range, while standard \DA{} greatly over-estimates $\Pkspec$ for the lower wavenumbers and, in some cases, even turns negative for the higher wavenumbers, thereby predicting unphysical upscale energy transfer that is absent in the ground truth data.

As for the energy spectra, the reconstructed energy fluxes
are seen to approach the ground truth as the initial fields are advanced in time
for both coarse-graining levels, as can be seen from the data shown in the middle and right panels of \autoref{fig:Pk3d_4DVarSR_varTunroll_var_time}. For $\nc = 8$, all methods reproduce the ground-truth fluxes rather well. However, for the strongest coarsening $\nc=16$ with data shown in subfigure (b), \DASR{} clearly out-performs the remaining techniques, providing very accurate reproductions of the ground truth flux 
even beyond the end of its assimilation window, where standard \DA{} and \SRdyn{} retain visible deviations from the ground truth.

%% file: sections/images/errL2uvw_vs_time.tex
\FPeval\TunrollNormA{round(0.5/\NHSteddy:1)}
\FPeval\TunrollNormB{round(1.0/\NHSteddy:1)}
\FPeval\TunrollNormC{round(1.5/\NHSteddy:1)}

\begin{figure}
  \centering
  \includegraphics[width=\linewidth]
  {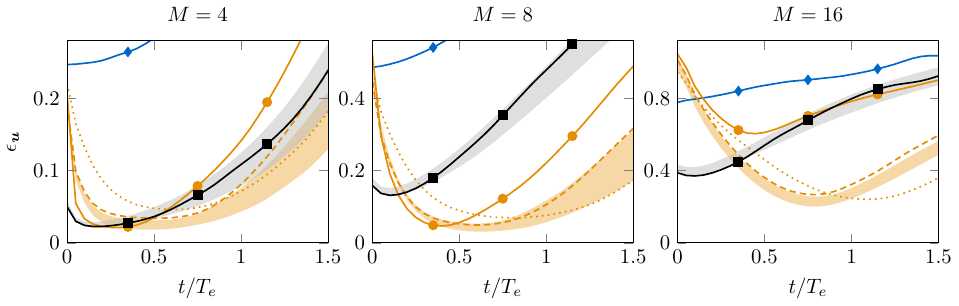}
  \caption
  { 
    Reconstruction error $\erruvw$ (equation~\eqref{eq:errL2uvw}) under time advancement
    for coarsening factors
    $\nc=4$,
    $\nc=8$ and
    $\nc=16$.
    Reconstructions shown are
    tri-cubic interpolation (blue),
    \SRdyn{} (black),
    \DA{} (orange, solid/dashed/dotted lines for 
    $\TunrollDA/\teddy =\TunrollNormA,\TunrollNormB,\TunrollNormC$ respectively).
    The grey (\SRdyn{}) and orange (\DA{}, \TunrollDA=0.9\teddy) shaded regions indicate the ensemble mean $\pm$ one standard deviation over 
    an extended set of trajectories with different, statistically independent initial conditions, while the solid lines are the representative initial condition discussed in much of \S\ref{sec:results}.
  }
\label{fig:errL2uvw_vs_time}
\end{figure}

%% file: sections/images/v_xyplanes_snaps_T0.tex
\begin{figure}
  \centering
  \includegraphics[width=\linewidth]
  {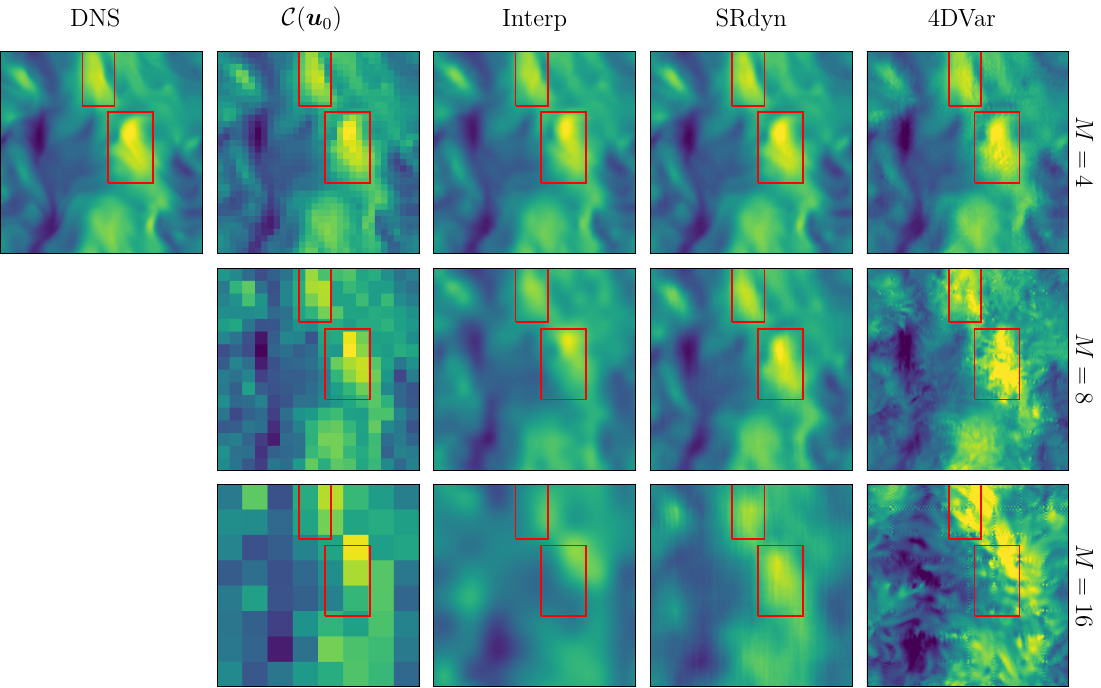}
  \caption{
    Comparison of the reconstructed velocity fields at $t=0$ with the high-resolution DNS snapshot that is to be reproduced, alongside the respective coarse-grained fields.
    Shown are slices of $v/\urms$ at $z=0$ for the different reconstruction procedures and coarsening factors $\nc\in\{4,8,16\}$, with colours ranging from blue (dark) to yellow (bright) in the interval $[-3,3]$.
    For the \DA{} runs, only the best performing case (in terms of the reconstruction error $\erruvw$, cf. equation~\eqref{eq:errL2uvw}) is shown for each coarsening factor.
    Red boxes highlight individual high-speed regions in the ground truth state and its reconstructions.
    }
\label{fig:v_xyplanes_snaps_T0}
\end{figure}

%% file: sections/images/uvw_omx_eps_pdf_var_time.tex
\begin{figure}
  \FPeval\TunrollNormA{round(0.5/\NHSteddy:1)}
  \FPeval\TunrollNormB{round(1.0/\NHSteddy:1)}
  \FPeval\TunrollNormC{round(1.5/\NHSteddy:1)}
  \centering
  \includegraphics[width=\linewidth]
  {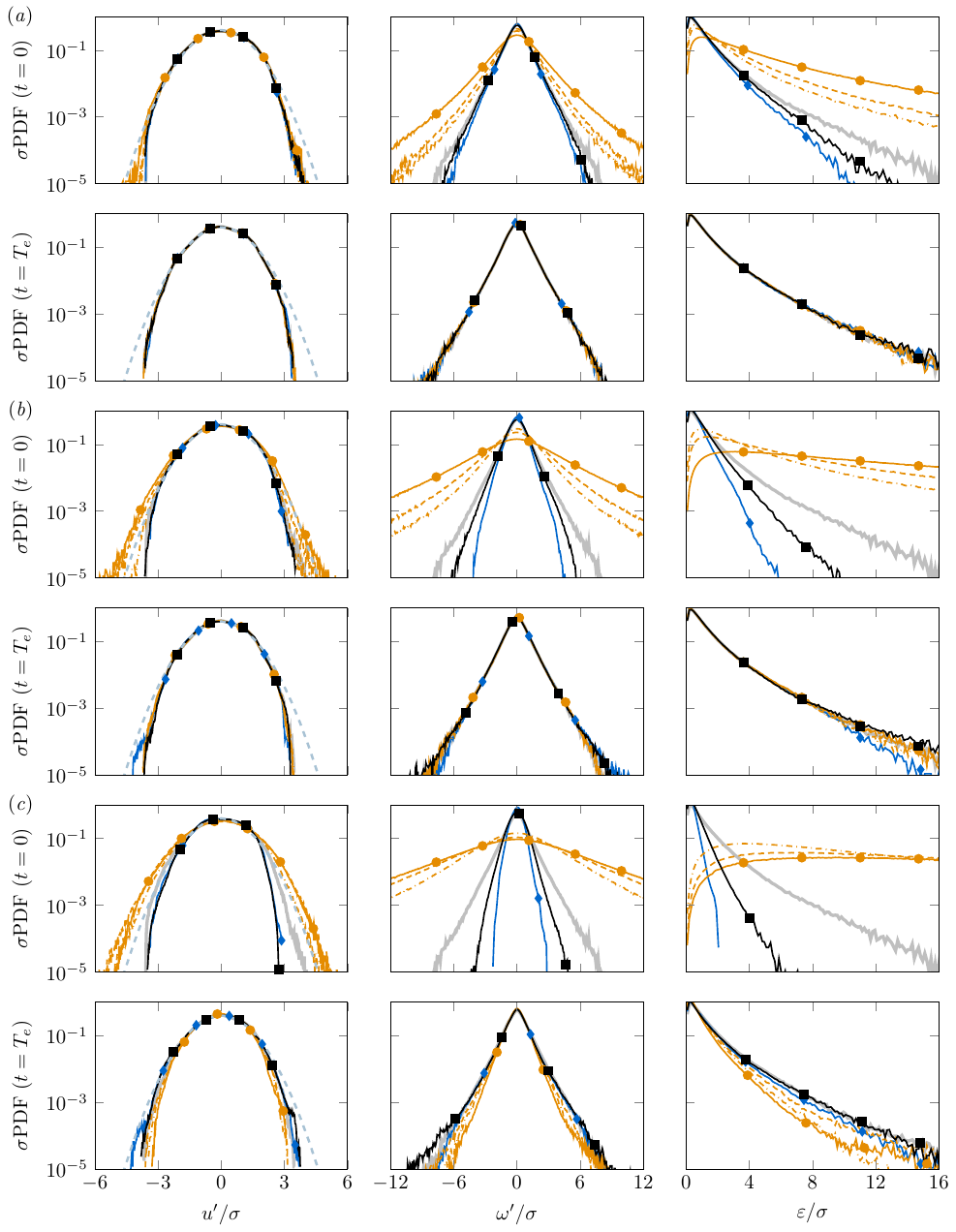}
  \caption{
    Standard probability density functions of
    (left) the velocity fluctuations $\uprime$ ($\sigma=\urms$),
    (middle) the vorticity fluctuations $\omprime$ ($\sigma=\omrms$) and
    (right) the local dissipation rate $\eps$ ($\sigma=\Diss$)
    for the reconstructed fields at $t=0$ (rows one, three and five)
    and $t=\teddy$ (rows two, four and six):
    (\textit{a}) $M=4$,
    (\textit{b}) $M=8$,
    (\textit{c}) $M=16$.
    Variables shown are the DNS ground truth (thick dark grey),
    tri-cubic interpolation (blue),
    \SRdyn{} (black),
    \DA{} (orange, line styles as before) and
    the standard normal distribution (light blue, dashed, shown for $\uprime$ only).
    }
\label{fig:uvw_omx_eps_pdf_var_time}
\end{figure}

%% file: sections/images/crosscorr_DNS_SRdyn_4DVar_var_time.tex
\begin{figure}
  \centering
  \includegraphics[width=\linewidth]
  {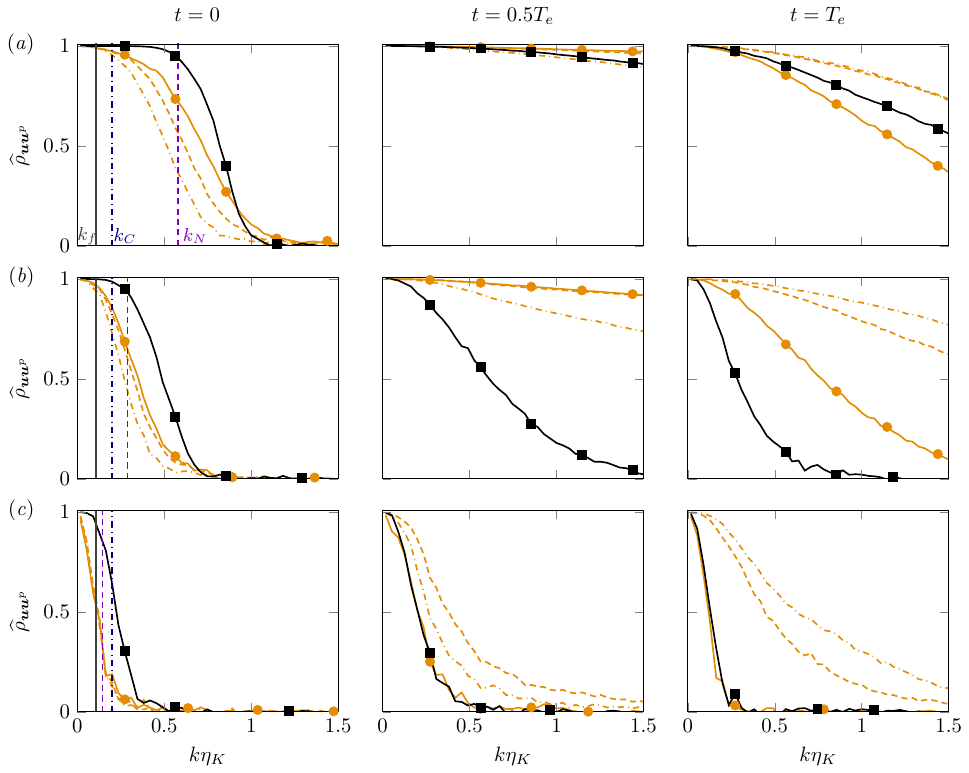}
  \caption{
    Normalised co-spectrum $\Crosscorr(k,t)$ between the ground truth $\uvec$ and the reconstructed field $\uvecpred$ for the \SRdyn{} (black) and \DA{} runs (orange, line styles as before) as a function of the wavenumber
    at $t/\teddy\in\{0,0.5,1\}$
    for coarsening factors
    (\textit{a}) $\nc=4$,
    (\textit{b}) $\nc=8$ and
    (\textit{c}) $\nc=16$.
    Vertical lines indicate
    the maximum forcing wavenumber $\kf=3$ (solid, grey),
    the Nyquist cutoff wavenumber $\kNyquist$ of the respective coarsening (dashed, purple) and
    the critical wavenumber $\kc=0.2\leta^{-1}$ (dash-dotted, blue),
    respectively.
    }
\label{fig:crosscorr_DNS_SRdyn_4DVar_var_time}
\end{figure}

%% file: sections/images/omy_xyplanes_snaps_DNS_SRdyn_4DVar_nc8_vs_time.tex
\FPeval\TunrollNormA{round(0.5/\NHSteddy:1)}
\FPeval\TunrollNormB{round(1.0/\NHSteddy:1)}
\FPeval\TunrollNormC{round(1.5/\NHSteddy:1)}

\begin{figure}
  \centering
  \includegraphics[width=\linewidth]
  {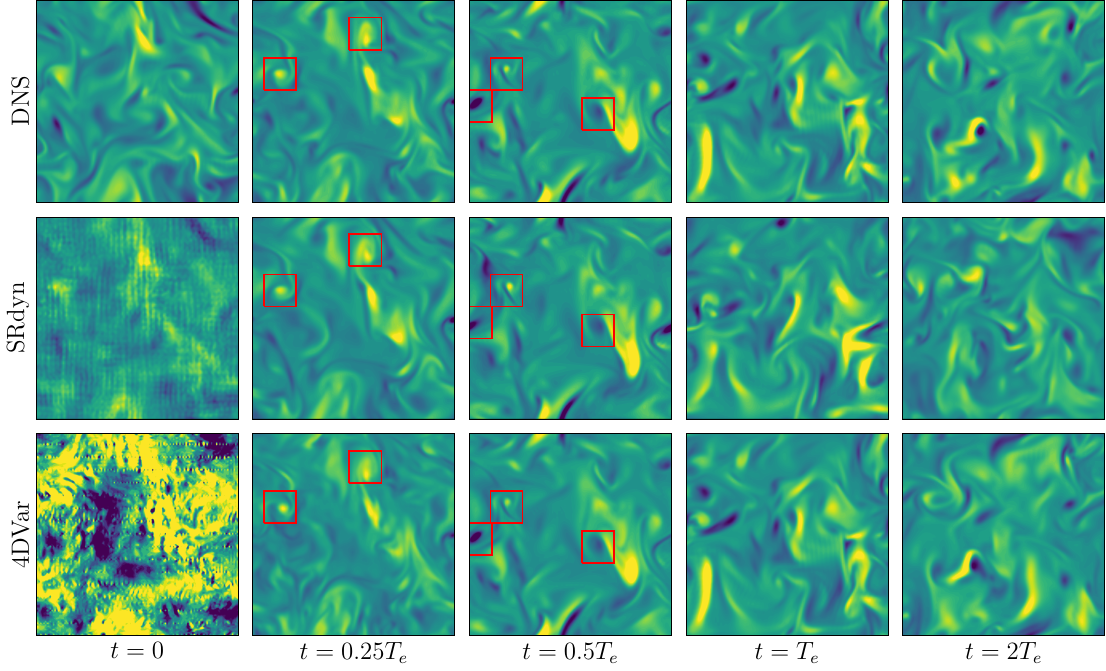}
  \caption{
    Comparison of the unrolled vorticity trajectory initialised with the reconstructed vorticity field $\omvecpredzero$
    for \SRdyn{} and \DA{} ($\TunrollDA\approx\TunrollNormC\teddy$) and a coarsening factor $\nc=8$.
    Shown are slices of $\omy/\omrms$ at $z=0$, with colouring as in \autoref{fig:omy_xyplanes_snaps_DNSlongtime}.
    Red boxes highlight individual structures in the ground truth state and their reconstructions.
    }
\label{fig:omy_xyplanes_snaps_DNS_SRdyn_4DVar_nc8_vs_time}
\end{figure}

%% file: sections/images/qhunt_3diso_DNS_SRdyn_4DVar_nc8_vs_time.tex
\begin{figure}
  \centering
  \includegraphics[width=\linewidth]
  {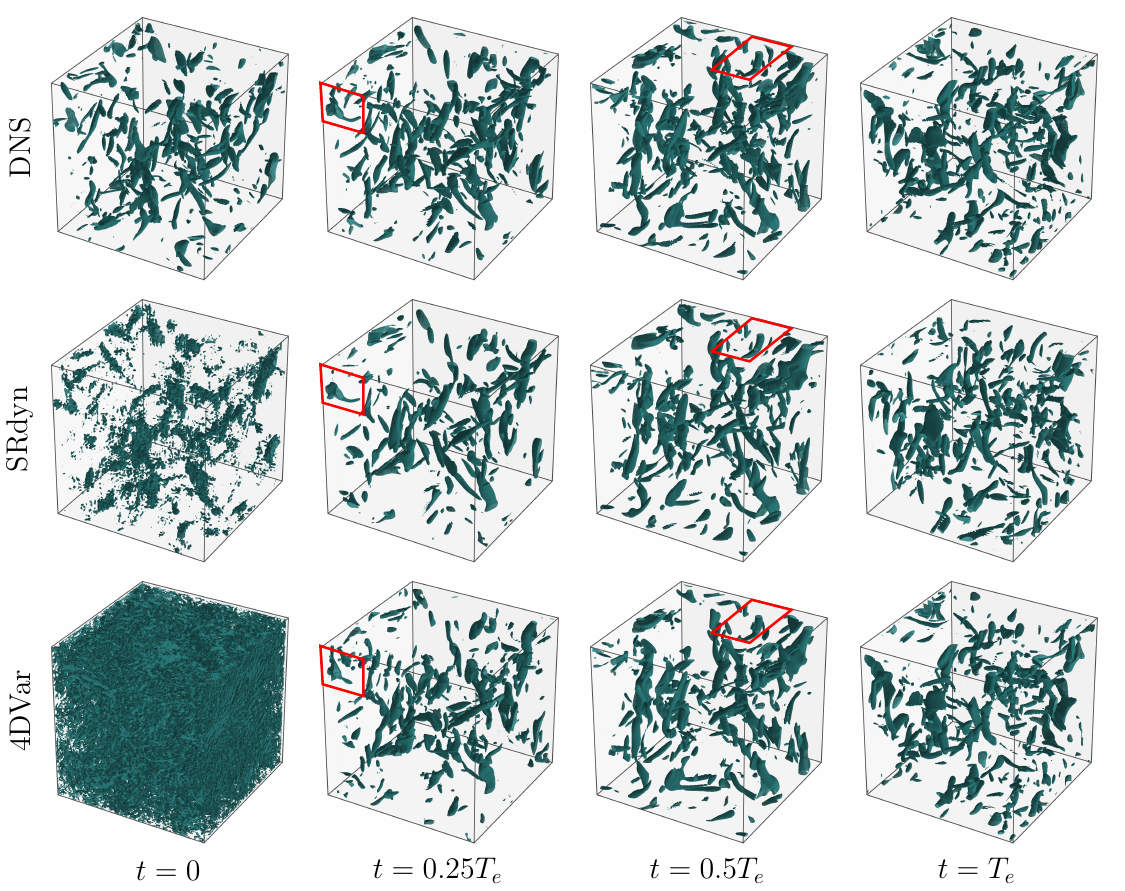}
  \caption{
    Evolution of vortical structures from state estimation at $M=8$, visualised using isosurfaces of the second invariant of the velocity gradient tensor $Q$ \citep{Hunt_al_1988} at a threshold $Q/\omrms^2 = 1$.
    The \DA{} results are obtained with an optimisation over a window of length $\TunrollDA = 1.3 \teddy$.
    Red boxes highlight individual vortical structures and their reconstructions.
    }
\label{fig:qhunt_3diso_DNS_SRdyn_4DVar_nc8_vs_time}
\end{figure}

%% file: sections/images/v_xyplanes_snaps_T0_DNS_SRdyn_4DVar_4DVarSR.tex
\begin{figure}
  \centering
  \includegraphics[width=\linewidth]
  {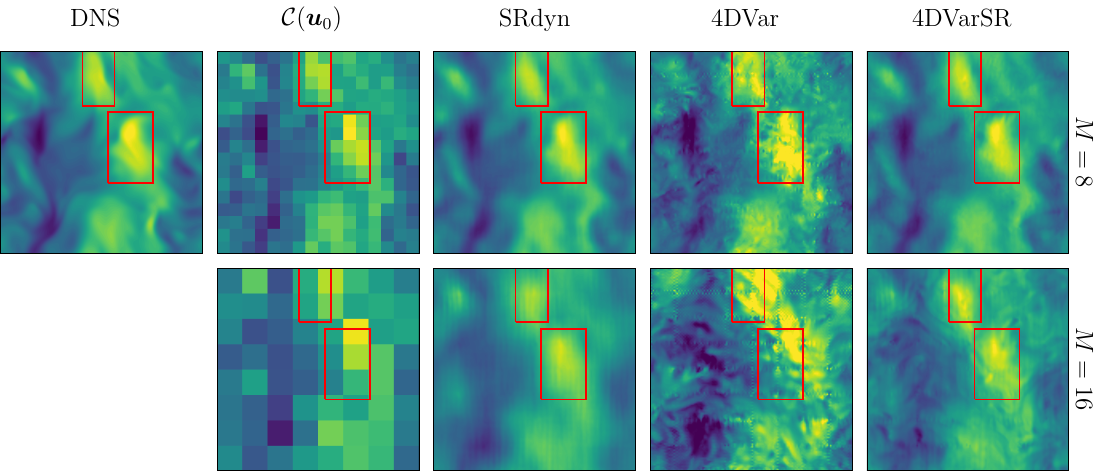}
  \caption{
    Comparison of the \DASR{}-reconstructed velocity fields at $t=0$ with the high-resolution DNS snapshot that is to be reproduced, the respective coarse-grained fields and those obtained with straight \SRdyn{} and \DA{}.
    Shown are slices of $v/\urms$ at $z=0$ for the different reconstruction procedures and coarsening factors $\nc\in\{8,16\}$, with colouring as in \autoref{fig:v_xyplanes_snaps_T0}.
    For the \DA{} and \DASR{} runs, only the best performing case is shown for each value of $\nc$.
    Position and size of the red boxes which highlight selected high-speed regions are identical to those in \autoref{fig:v_xyplanes_snaps_T0}.
    }
\label{fig:v_xyplanes_snaps_T0_DNS_SRdyn_4DVar_4DVarSR}
\end{figure}

%% file: sections/images/uvw_omx_eps_pdf_DNS_4DVarSR_nc8_nc16_var_time.tex
\begin{figure}
  \centering
  \includegraphics[width=\linewidth]
  {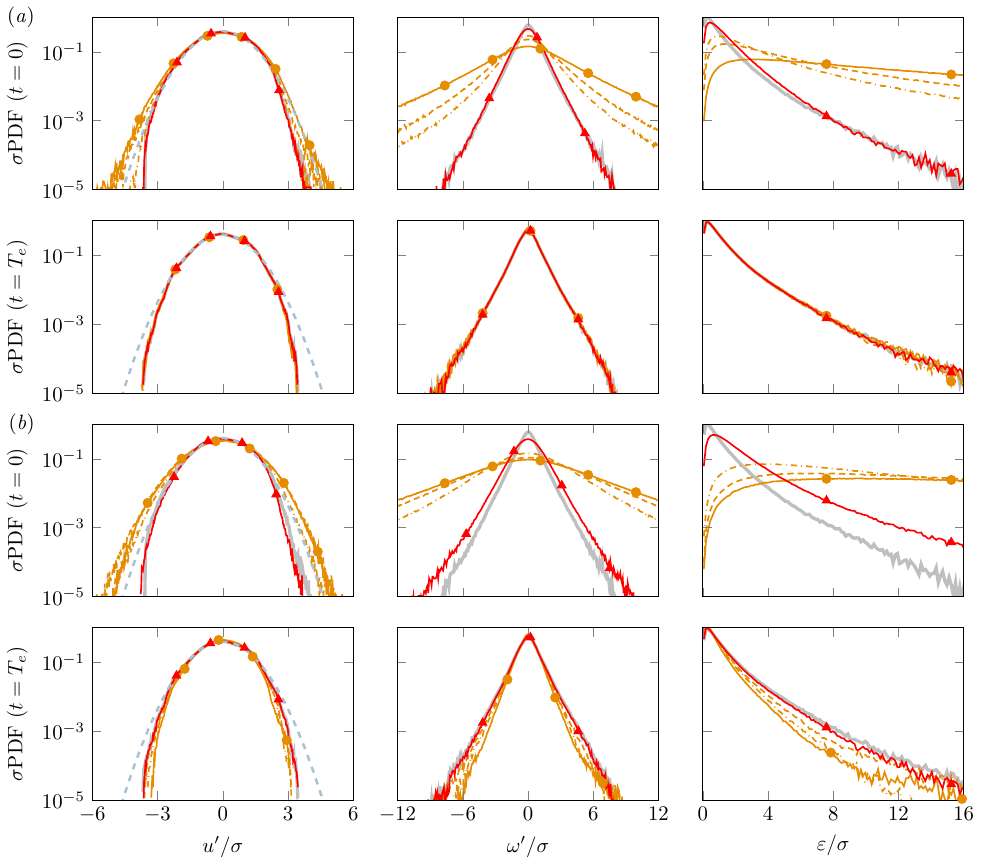}
  \caption{
    Standard probability density functions of
    (left) the velocity fluctuations $\uprime$ ($\sigma=\urms$),
    (middle) the vorticity fluctuations $\omprime$ ($\sigma=\omrms$) and
    (right) the local dissipation rate $\eps$ ($\sigma=\Diss$)
    for the reconstructed fields at $t=0$ (rows one and three)
    and $t=\teddy$ (rows two and four):
    (\textit{a}) $\nc=8$,
    (\textit{b}) $\nc=16$.
    Variables shown are the DNS ground truth (thick dark grey),
    \DA{} (orange, line styles as before),
    the best-performing \DASR{} run (red) and
    the standard normal distribution (light blue, dashed, shown for $\uprime$ only).
    }
\label{fig:uvw_omx_eps_pdf_DNS_4DVarSR_nc8_nc16_var_time}
\end{figure}

%% file: sections/images/errL2uvw_4DVarSR_varTunroll_vs_time.tex
\begin{figure}
  \centering
  \includegraphics[width=\linewidth]
  {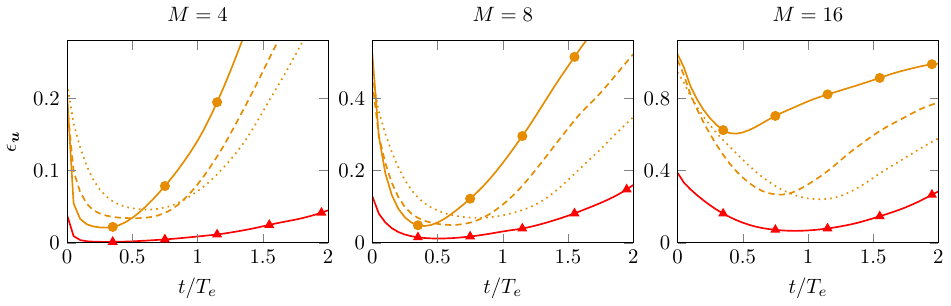}
  \caption
  {
    Comparison of the model predictions under time advancement in terms
    of the reconstruction error~$\erruvw$
    for coarsening factors
    $\nc=4$,
    $\nc=8$ and
    $\nc=16$.
    Reconstructions shown are
    \DA{} (orange, line styles as before) and
    the best-performing \DASR{} run (red).
  }
\label{fig:errL2uvw_4DVarSR_varTunroll_vs_time}
\end{figure}

%% file: sections/images/crosscorr_DNS_4DVar_4DVarSR_var_time.tex
\begin{figure}
  \centering
  \includegraphics[width=\linewidth]
  {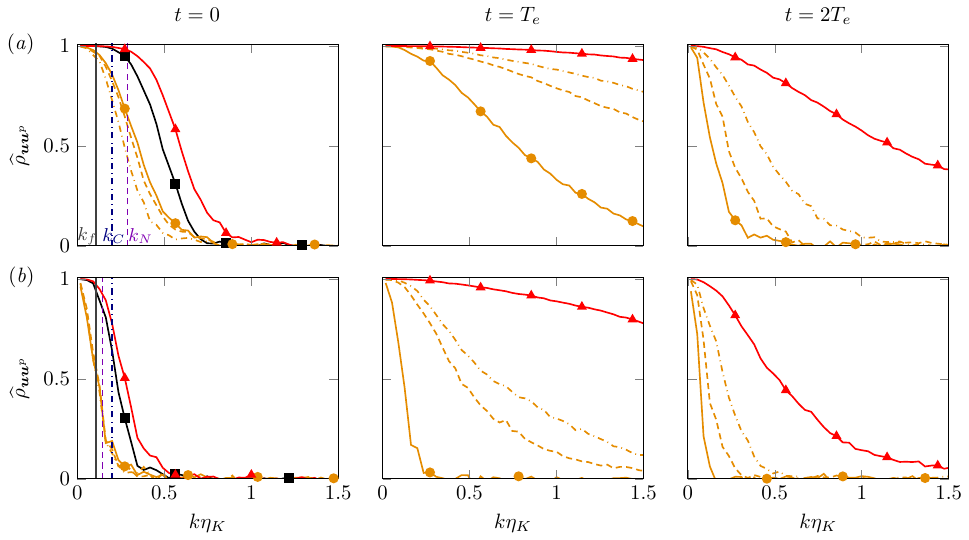}
  \caption{
    Normalised co-spectrum $\Crosscorr(k,t)$ between the ground truth $\uvec$ and the reconstructed field $\uvecpred$ at $t/\teddy\in\{0,1,2\}$
    for coarsening factors
    (\textit{a}) $\nc=8$ and
    (\textit{b}) $\nc=16$.
    Shown here are
    \DA{} (orange, line styles as before) and
    the best-performing \DASR{} run (red).
    At $t=0$, the \SRdyn{}-based reconstruction (black) is shown for comparison as well.
    Vertical lines are as in \autoref{fig:crosscorr_DNS_SRdyn_4DVar_var_time}.
    }
\label{fig:crosscorr_DNS_4DVar_4DVarSR_var_time}
\end{figure}

%% file: sections/images/omy_xyplanes_snaps_DNS_4DVar_4DVarSR_nc16_vs_time.tex
\FPeval\TunrollNormA{round(0.5/\NHSteddy:1)}
\FPeval\TunrollNormB{round(1.0/\NHSteddy:1)}
\FPeval\TunrollNormC{round(1.5/\NHSteddy:1)}

\begin{figure}
  \centering
  \includegraphics[width=\linewidth]
  {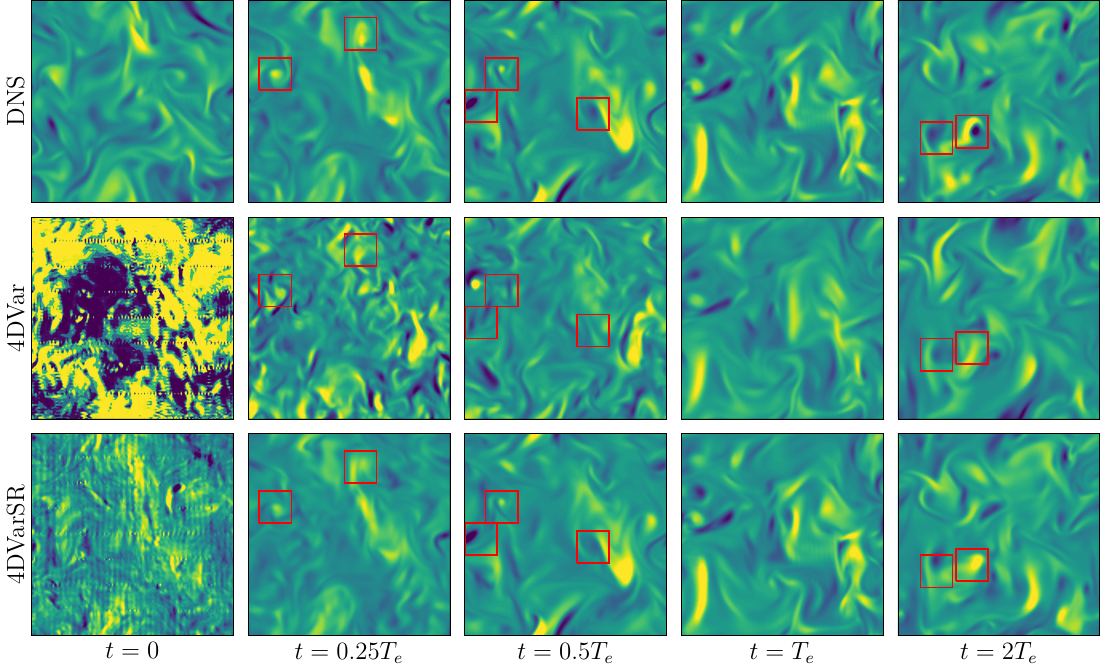}
  \caption{
    Comparison of the unrolled vorticity trajectory for `standard' 4DVar and \DASR{}
    (both $\TunrollDA\approx\TunrollNormC\teddy$) and a coarsening factor $\nc=16$.
    Shown are slices of $\omy/\omrms$ at $z=0$, with colouring as in \autoref{fig:omy_xyplanes_snaps_DNSlongtime}.
    Red boxes highlight individual structures in the ground truth state and their reconstructions.
    }
\label{fig:omy_xyplanes_snaps_DNS_4DVar_4DVarSR_nc16_vs_time}
\end{figure}

%% file: sections/images/omy_xyplanes_errLocal_DNS_4DVar_4DVarSR_nc16_vs_time.tex
\FPeval\TunrollNormA{round(0.5/\NHSteddy:1)}
\FPeval\TunrollNormB{round(1.0/\NHSteddy:1)}
\FPeval\TunrollNormC{round(1.5/\NHSteddy:1)}

\begin{figure}
  \centering
  \includegraphics[width=\linewidth]
  {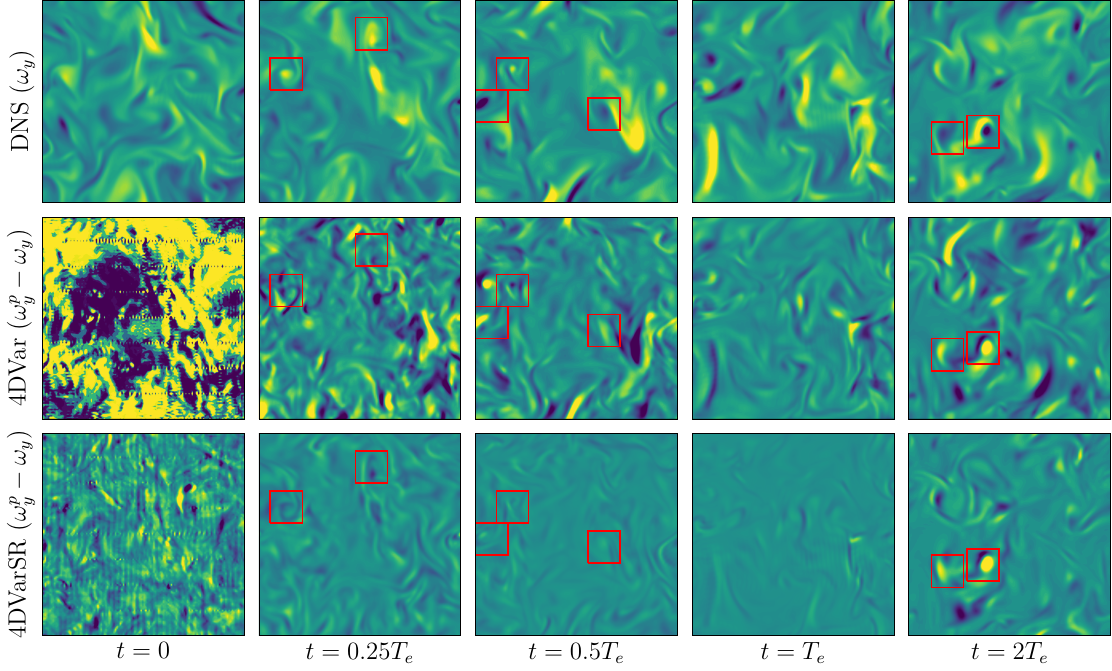}
  \caption{
    Comparison of the local reconstruction error $(\omypred-\omy)/\omrms$ at $z=0$ along the unrolled vorticity trajectories for 4DVar and \DASR{}
    (both $\TunrollDA\approx\TunrollNormC\teddy$) and a coarsening factor $\nc=16$.
    For the sake of comparison, `full' snapshots $\omy/\omrms$ from the ground truth DNS trajectory are included in the top row.
    Colours range from blue (dark) to yellow (light) in the interval $[-4,4]$.
    Red boxes highlight individual structures in the ground truth state and the respective local deviations in the reconstructed states.
    }
\label{fig:omy_xyplanes_errLocal_DNS_4DVar_4DVarSR_nc16_vs_time}
\end{figure}

%% file: sections/images/ek3d_4DVarSR_varTunroll_var_time.tex
\begin{figure}
  \centering
  \includegraphics[width=\linewidth]
  {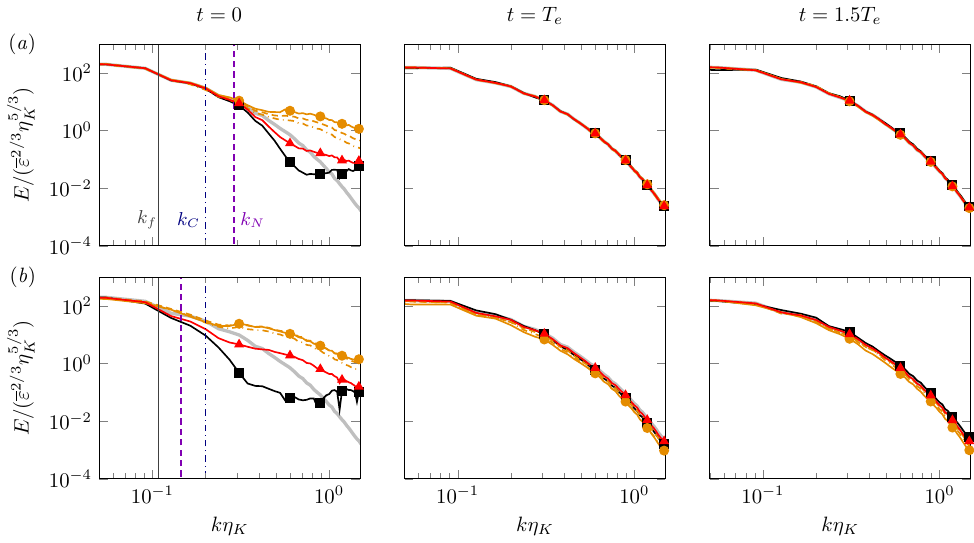}
  \caption{
    Instantaneous normalised energy spectra 
    as a function of the wavenumber for snapshots at $t/\teddy\in\{0,1,1.5\}$
    and coarsening factors
    (\textit{a}) $\nc=8$ and
    (\textit{b}) $\nc=16$.
    Shown here are the DNS ground truth (thick grey),
    \SRdyn{} (black),
    standard \DA{} (orange, line styles as before) and
    the best-performing \DASR{} run (red).
    Vertical lines are as in \autoref{fig:crosscorr_DNS_SRdyn_4DVar_var_time}.
    }
\label{fig:ek3d_4DVarSR_varTunroll_var_time}
\end{figure}

%% file: sections/images/Pk3d_4DVarSR_varTunroll_var_time.tex
\begin{figure}
  \centering
  \includegraphics[width=\linewidth]
  {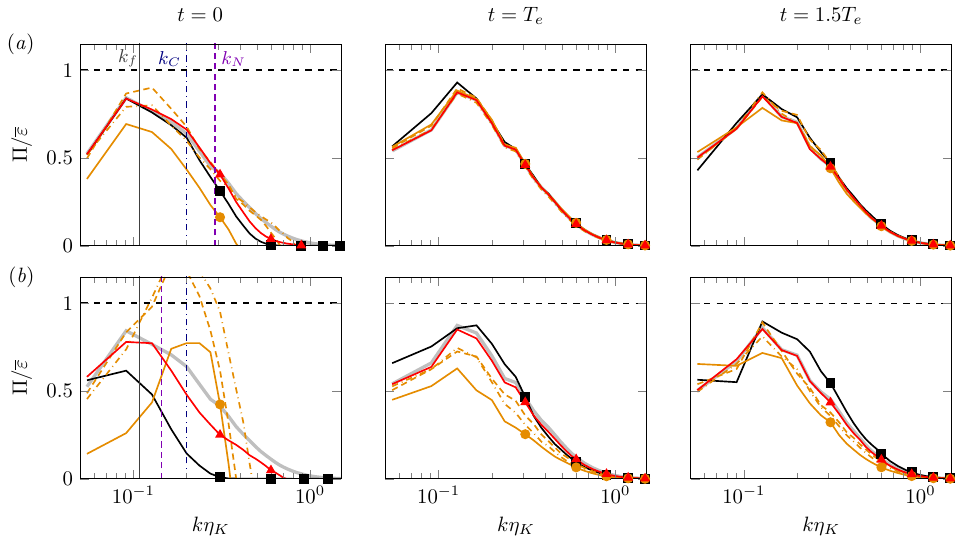}
  \caption{
    Instantaneous normalised energy fluxes, $\Pkspec/\Diss$, as a function of the wavenumber for snapshots at $t/\teddy\in\{0,1,1.5\}$
    and coarsening factors
    (\textit{a}) $\nc=8$ and
    (\textit{b}) $\nc=16$.
    Shown here are the DNS ground truth (thick grey), 
    \SRdyn{} (black), 
    standard \DA{} (orange, line styles as before) and 
    the best-performing \DASR{} run (red).
    Vertical lines are as in \autoref{fig:crosscorr_DNS_SRdyn_4DVar_var_time},
    and the horizontal line marks $\Pkspec=\Diss$.
    }
\label{fig:Pk3d_4DVarSR_varTunroll_var_time}
\end{figure}

%% file: sections/conclusion.tex
In this paper, we have presented a trajectory-based super-resolution technique for state estimation in triply-periodic homogeneous isotropic turbulence.
The training process is inspired by variational data assimilation in that the network reconstructions are unrolled in time and trained against time series of coarse-grained velocity fields only
-- network training does not require a library of high-resolution snapshots \citep{Page_2025}.
A fully-differentiable three-dimensional pseudo-spectral DNS code was used to advance the network outputs in time, which enables gradient descent on a loss involving entire trajectories.

The trained networks generate robust reconstructions of previously unseen high-resolution velocity snapshots from coarse-grained snapshots of homogeneous isotropic turbulence at $\Relambda\approx70$.
Notably, the initial pointwise reconstruction errors achieved by the super-resolution approach represent a $50\%$ or more reduction to classical \DA{}, which is an optimisation to match low-resolution observations along the specific target trajectory.
Unrolling the reconstructed fields in time, \DA{} successively improves in its predictive accuracy, reaching a comparable precision to super-resolution after $0.2$ to $0.5$ large-eddy times (depending on the coarse-graining level).
The good performance of super-resolution at early times can presumably be attributed to the networks having learned to construct physically realistic representations of the solution manifold from exposure to a large variety of coarse trajectories during the training.
In a second step, we explored an alterative initialisation for the \DA{} algorithm using the super-resolution neural networks trained on coarse trajectories.
The initialization with a super-resolved field is highly beneficial for the performance of \DA{}, with the hybrid approach providing more accurate predictions of the ground truth trajectory over the entire assimilation window. Notably, robust predictions are achieved even for a coarse-graining levels clearly above the synchronisation limit $\lc$ at which data assimilation usually struggles.

The results showcase the versatility of the new super-resolution technique and highlight the benefits of including fundamental physical principles in the network training process.
Naturally, the observed robust predictions achieved by trajectory-based super-resolution raise the question how the models will perform at higher Reynolds numbers, for which turbulence exhibits a pronounced inertial range over a wide range of scales.
However, the significant computational resources that are required for such an undertaking should not be disregarded -- as a reference, training of the current networks at a target resolution of $128^3$ grid points took roughly three hours per epoch on a 140GB NVIDIA H200 card.
Each training epoch involves $O(10^3)$ individual data assimilations performed in parallel over minibatches of size $16$, so the full training process is equivalent to performing more than $\num{11000}$ individual assimilations.
The advantage though is that once the model is trained, reconstructing a high-resolution field reduces to a single function call, while data assimilation has to be redone for each individual trajectory.
Conversely, performing data assimilation starting from a super-resolved initial guess does not come with additional costs compared to a traditional initialization once the model has been trained.
Given the expected high computational costs when targetting significantly higher Reynolds numbers, it might be beneficial to switch to a model architecture that specifically takes into account the multi-scale structure of the turbulent fields \citep{Fukami_Fukagata_Taira_2019}. Also, recent experience in the low-data limit might help to maintain the reconstruction accuracy when training is performed on smaller datasets \citep{Fukami_Taira_2024}.

Moreover, we see great potential in generalizing the concept of a trajectory-based learning to situations for which observations are only available at arbitrarily distributed probing or sensor locations as in many experiments.
While the `classical' CNN architectures used here cannot directly deal with such irregularly distributed input data, other model architectures (e.g. Graph Neural Networks or point-cloud-based convolutions) can easily be incorporated into the training procedure outlined in this work.

%% file: sections/funding.tex
MW thanks the Karlsruhe House of Young Scientists (KHYS) for their financial support of his research stay at the University of Edinburgh.
JP acknowledges support from a UKRI Frontier Guarantee Grant EP/Y004094/1. 
Computational resources were kindly provided by the Edinburgh International Data Facility (EIDF), the Data-Driven Innovation Programme at the University of Edinburgh and on the supercomputer bwUniCluster funded by the Ministry of Science, Research and the Arts Baden-W{\"u}rttemberg and the Universities of the State of Baden-W{\"u}rttemberg.
The computer resources, technical expertise and assistance provided from the technical staff are gratefully acknowledged.

%% file: sections/appendixKF.tex
%
%
\input{sections/tables/KF_parameter_tables.tex}
%
As described in \S\ref{subsec:SRdyn}, the \DA{} procedure proposed by \citet{Li_Zhang_Dong_Abdullah_2020} operates in spectral space and is therefore provided with the ground truth Fourier modes at wavenumbers $0\leq k\leq k_O$ as the measurements for comparison.
To quantify the difference in performance between this purely spectral approach and the physical-space \DA{} performed here, we present here some additional physical-space \DA{} runs that have been performed for the unidirectional Kolmogorov forcing and Reynolds number $\Relambda\approx 75$ to match the flow configuration considered in \citet{Li_Zhang_Dong_Abdullah_2020}.
The relevant physical and numerical parameters of these simulations are summarised in \autoref{tab:KF_param_phys_num}.
For consistency with the data presented in \citet{Li_Zhang_Dong_Abdullah_2020}, time is henceforth measured in terms of a forcing-related time-scale $\teddyLi \coloneq \urms/\forceampl$, where $\forceampl$ is the forcing amplitude in the Kolmogorov forcing
\begin{equation}
  \fvecext^*=(0,\forceampl \cos(2\pi\kf x^*/L*),0)^T
\end{equation}
before non-dimensionalisation. Here, the forcing wavenumber is set at $\kf=1$ and $\teddyLi \approx 0.76\teddy$ for the chosen parameter point.

\input{sections/images/KF_crosscorr_DNS_4DVar_var_time.tex}
%
The mode-wise reconstruction error $\Crosscorr$ is shown in \autoref{fig:KF_crosscorr_DNS_4DVar_var_time} for coarse-graining levels $\nc\in\{8,16,32\}$ and two different assimilation windows with length $\TunrollDA/\teddyLi\in\{0.5,1\}$.
Where available, corresponding data points from \citet{Li_Zhang_Dong_Abdullah_2020} (see their Figure 7c,e) with matching cutoff wavenumber for the spectral low pass filter $k_O=\kNyquist$ are included in red.
The largest differences between the Fourier-based \DA{} results of \citet{Li_Zhang_Dong_Abdullah_2020} and those obtained in the current work based on coarse-grained velocity snapshots are seen at $\nc=16$, where the former leads to a more accurate reproduction of the small scales.
At the weakest coarse-graining $\nc=8$, on the other hand, both methods lead to almost identical results.
Finally, for $\nc=32$, relevant deviations are restricted to the first few Fourier modes; both methods struggle to reproduce the modes at $k>0.25/\leta$.

%% file: sections/tables/KF_parameter_tables.tex
\FPeval\Lbox{2*\FPpi}
\FPeval\LeddyoL{\KFLeddy/\Lbox}
\FPeval\lambdaoL{\KFlambdaTaylor/\Lbox}
\FPeval\Loleta{\Lbox/\KFeta}
\FPeval\lambdaoleta{\KFlambdaTaylor/\KFeta}
\FPeval\Leddyoeta{\KFLeddy/\KFeta}

\FPeval\kfEta{3 * \KFeta}
\FPeval\kmaxEta{trunc(trunc(2./3.*128:0)/2.:0) * \KFeta}
\FPeval\kNyquistetaOne{128/(2*4)*\KFeta}
\FPeval\kNyquistetaTwo{128/(2*8)*\KFeta}
\FPeval\kNyquistetaThree{128/(2*16)*\KFeta}
\FPeval\kNyquistetaFour{128/(2*32)*\KFeta}
\FPeval\dxEta{\Lbox/128 / \KFeta}
\FPeval\dxcEtaOne{\Lbox/128 * 4 / \KFeta}
\FPeval\dxcEtaTwo{\Lbox/128 * 8 / \KFeta}
\FPeval\dxcEtaThree{\Lbox/128 * 16 / \KFeta}
\FPeval\dxcEtaFour{\Lbox/128 * 32 / \KFeta}
\FPeval\dxlc{\dxEta/(5*\FPpi)}
\FPeval\dxlcOne{\dxcEtaOne/(5*\FPpi)}
\FPeval\dxlcTwo{\dxcEtaTwo/(5*\FPpi)}
\FPeval\dxlcThree{\dxcEtaThree/(5*\FPpi)}
\FPeval\dxlcFour{\dxcEtaFour/(5*\FPpi)}
\FPeval\lcdx{1/\dxlc}

\nprounddigits{2}

\begin{table}
  \begin{center}
    \hrulefill \\
    \hfill \\
    \begin{tabular}{c c H S S S S S S S S}
      \multicolumn{1}{c}{Setting}&
      \multicolumn{1}{c}{$\Ni$}&
      \multicolumn{1}{H}{$\kf\leta$}&
      \multicolumn{1}{c}{$k_{(m,N)}\leta$}&
      \multicolumn{1}{c}{$\deltax_{(c)}/\leta$}&
      \multicolumn{1}{c}{$\deltax_{(c)}/\lc$}&
      \multicolumn{1}{c}{$\ReLeddy$}&
      \multicolumn{1}{c}{$\Relambda$}&
      \multicolumn{1}{c}{$\Lxyz/\leta$}&
      \multicolumn{1}{c}{$\Leddy/\leta$}&
      \multicolumn{1}{c}{$\lambdaTaylor/\leta$}\\[2.ex]
      \DNS & $128$ & $\np{\kfEta}$ & $\np{\kmaxEta}$ & $\np{\dxEta}$ & $\np{\dxlc}$ &
      $\np{\KFReLeddy}$ & $\np{\KFRelambda}$ &
      $\np{\Loleta}$ & $\np{\Leddyoeta}$ &
      $\np{\lambdaoleta}$ \\[2.0ex] 
      $\nc=\,4$ & $32$ & $\np{\kfEta}$ & \np{\kNyquistetaOne} & $\np{\dxcEtaOne}$ & $\np{\dxlcOne}$ &
      & &
      & &
      \\[1.ex]
      $\nc=\,8$ & $16$ & $\np{\kfEta}$ & \np{\kNyquistetaTwo} & $\np{\dxcEtaTwo}$ & $\np{\dxlcTwo}$ &
      & &
      & &
      \\[1.ex]
      $\nc=16$ & $8$ & $\np{\kfEta}$ & \np{\kNyquistetaThree} & $\np{\dxcEtaThree}$ & $\np{\dxlcThree}$ &
      & &
      & &
      \\[1.ex]
      $\nc=32$ & $4$ & $\np{\kfEta}$ & \np{\kNyquistetaFour} & $\np{\dxcEtaFour}$ & $\np{\dxlcFour}$ &
      & &
      & &
      \\[1.ex]
    \end{tabular}
    \caption{
      Physical and numerical parameters of the DNS database for the Kolmogorov flow setting studied by \citet{Li_Zhang_Dong_Abdullah_2020},
      together with information on the low-resolution fields for a coarsening factor $\nc$.
      Definitions match those in \autoref{tab:param_phys_num}.
    }
    \label{tab:KF_param_phys_num}
  \end{center}
  \vspace{-2ex}
  \hrulefill
\end{table}

%% file: sections/images/KF_crosscorr_DNS_4DVar_var_time.tex
\begin{figure}
  \centering
  \includegraphics[width=\linewidth]
  {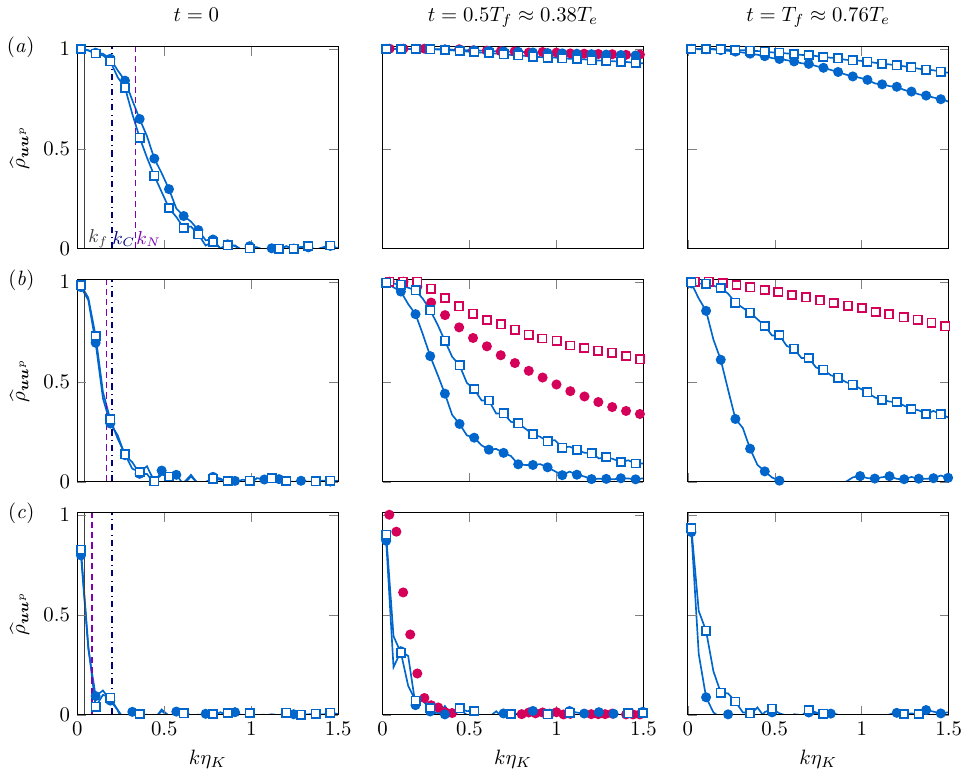}
  \caption{
    Normalised co-spectrum $\Crosscorr(k,t)$ between the ground truth $\uvec$ and the \DA{}-based reconstruction $\uvecpred$ at $t/\teddyLi\in\{0,0.5,1\}$ for the Kolmogorov flow setting studied by \citet{Li_Zhang_Dong_Abdullah_2020};
    coarsening factors are
    (\textit{a}) $\nc=8$ ($\kNyquist=8$),
    (\textit{b}) $\nc=16$ ($\kNyquist=4$) and
    (\textit{c}) $\nc=32$ ($\kNyquist=2$).
    Results shown are the current \DA{} runs (blue) and
    data from \citet{Li_Zhang_Dong_Abdullah_2020} for $k_O=\kNyquist$ (red) for
    $\TunrollDA=0.5\teddyLi$ (solid circles) and
    $\TunrollDA=\teddyLi$ (open squares).
    Note that the results of \citet{Li_Zhang_Dong_Abdullah_2020} are obtained as ensemble averages over $O(1)$ individual realizations and are not available for all times.
    Vertical lines indicate
    the forcing wavenumber $\kf=1$ (solid, grey),
    the Nyquist cutoff wavenumber $\kNyquist$ of the respective coarsening (dashed, purple) and
    the critical wavenumber $\kc=0.2\leta^{-1}$ (dash-dotted, blue),
    respectively.
    }
\label{fig:KF_crosscorr_DNS_4DVar_var_time}
\end{figure}

%% file: main.bbl
\begin{thebibliography}{62}
\providecommand{\natexlab}[1]{#1}
\providecommand{\url}[1]{\texttt{#1}}
\expandafter\ifx\csname urlstyle\endcsname\relax
  \providecommand{\doi}[1]{doi: #1}\else
  \providecommand{\doi}{doi: \begingroup \urlstyle{rm}\Url}\fi

\bibitem[Alhashim et~al.(2025)Alhashim, Hausknecht, and Brenner]{Alhashim2025}
Mohammed~G. Alhashim, Kaylie Hausknecht, and Michael~P. Brenner.
\newblock Control of flow behavior in complex fluids using automatic
  differentiation.
\newblock \emph{Proc. Natl. Acad. Sci.},
  \href{http://dx.doi.org/10.1073/pnas.2403644122}{122\penalty0 (8)}, 2025.
\newblock ISSN 1091-6490.

\bibitem[Bannister(2017)]{Bannister_2017}
R.~N. Bannister.
\newblock A review of operational methods of variational and
  ensemble-variational data assimilation.
\newblock \emph{Q. J. R. Meteorol. Soc.},
  \href{http://dx.doi.org/10.1002/qj.2982}{143\penalty0 (703):\penalty0
  607--633}, 2017.

\bibitem[Barth{\'e}l{\'e}my et~al.(2022)Barth{\'e}l{\'e}my, Brajard, Bertino,
  and Counillon]{Barthelemy_Brajard_Bertino_Counillon_2022}
S.~Barth{\'e}l{\'e}my, J.~Brajard, L.~Bertino, and F.~Counillon.
\newblock Super-resolution data assimilation.
\newblock \emph{Ocean Dyn.},
  \href{http://dx.doi.org/10.1007/s10236-022-01523-x}{72:\penalty0 661--678},
  2022.

\bibitem[Baydin et~al.(2018)Baydin, Pearlmutter, Radul, and
  Siskind]{Baydin_Pearlmutter_Radul_Siskind_2018}
A.~G. Baydin, B.~A. Pearlmutter, A.~A. Radul, and J.~M. Siskind.
\newblock Automatic differentiation in machine learning: a survey.
\newblock \emph{J.\ Machine Learning Res.}, 18\penalty0 (153):\penalty0 1--43,
  2018.

\bibitem[Bewley and Protas(2004)]{Bewley_Protas_2004}
T.~R. Bewley and B.~Protas.
\newblock Skin friction and pressure: the ``footprints'' of turbulence.
\newblock \emph{Phys.\ D},
  \href{http://dx.doi.org/10.1016/j.physd.2004.02.008}{196\penalty0
  (1):\penalty0 28--44}, 2004.

\bibitem[Boffetta and Musacchio(2017)]{Boffetta_Musacchio_2017}
G.~Boffetta and S.~Musacchio.
\newblock Chaos and predictability of homogeneous-isotropic turbulence.
\newblock \emph{Phys. Rev. Lett.},
  \href{http://dx.doi.org/10.1103/PhysRevLett.119.054102}{119:\penalty0
  054102}, 2017.

\bibitem[Bradbury et~al.(2018)Bradbury, Frostig, Hawkins, Johnson, Leary,
  Maclaurin, Necula, Paszke, Vander{P}las, Wanderman-{M}ilne, and
  Zhang]{Jax_github}
J.~Bradbury, R.~Frostig, P.~Hawkins, M.~J. Johnson, C.~Leary, D.~Maclaurin,
  G.~Necula, A.~Paszke, J.~Vander{P}las, S.~Wanderman-{M}ilne, and Q.~Zhang.
\newblock {JAX}: composable transformations of {P}ython+{N}um{P}y programs,
  2018.

\bibitem[Brenner et~al.(2019)Brenner, Eldredge, and
  Freund]{Brenner_Eldredge_Freund_2019}
M.~P. Brenner, J.~D. Eldredge, and J.~B. Freund.
\newblock Perspective on machine learning for advancing fluid mechanics.
\newblock \emph{Phys. Rev. Fluids},
  \href{http://dx.doi.org/10.1103/PhysRevFluids.4.100501}{4:\penalty0 100501},
  2019.

\bibitem[Brunton et~al.(2020)Brunton, Noack, and
  Koumoutsakos]{Brunton_Noack_Koumoutsakos_2020}
S.~L. Brunton, B.~R. Noack, and P.~Koumoutsakos.
\newblock Machine learning for fluid mechanics.
\newblock \emph{Ann.\ Rev.\ Fluid Mech.},
  \href{http://dx.doi.org/10.1146/annurev-fluid-010719-060214}{52:\penalty0
  477--508}, 2020.

\bibitem[Buchta and Zaki(2021)]{Buchta_Zaki_2021}
D.~A. Buchta and T.~A. Zaki.
\newblock Observation-infused simulations of high-speed boundary-layer
  transition.
\newblock \emph{J.\ Fluid Mech.},
  \href{http://dx.doi.org/10.1017/jfm.2021.172}{916:\penalty0 A44}, 2021.

\bibitem[Chevalier et~al.(2006)Chevalier, H{\oe}pffner, Bewley, and
  Henningson]{Chevalier_Hoepffner_Bewley_Henningson_2006}
M.~Chevalier, J.~H{\oe}pffner, T.~R. Bewley, and D.~S. Henningson.
\newblock State estimation in wall-bounded flow systems. part 2. turbulent
  flows.
\newblock \emph{J.\ Fluid Mech.},
  \href{http://dx.doi.org/10.1017/S0022112005008578}{552:\penalty0 167–187},
  2006.

\bibitem[Chollet et~al.(2015)]{Chollet_2015}
Fran\c{c}ois Chollet et~al.
\newblock Keras.
\newblock \url{https://keras.io}, 2015.

\bibitem[Colburn et~al.(2011)Colburn, Cessna, and
  Bewley]{Colburn_Cessna_Bewley_2011}
C.~H. Colburn, J.~B. Cessna, and T.~R. Bewley.
\newblock State estimation in wall-bounded flow systems. part 3. the ensemble
  kalman filter.
\newblock \emph{J.\ Fluid Mech.},
  \href{http://dx.doi.org/10.1017/jfm.2011.222}{682:\penalty0 289–303}, 2011.

\bibitem[Di~Leoni et~al.(2020)Di~Leoni, Mazzino, and
  Biferale]{DiLeoni_Mazzino_Biferale_2020}
P.~C. Di~Leoni, A.~Mazzino, and L.~Biferale.
\newblock Synchronization to big data: Nudging the {N}avier-{S}tokes equations
  for data assimilation of turbulent flows.
\newblock \emph{Phys. Rev. X},
  \href{http://dx.doi.org/10.1103/PhysRevX.10.011023}{10:\penalty0 011023},
  2020.

\bibitem[Dong et~al.(2016)Dong, Loy, He, and Tang]{Dong_Loy_He_Tang_2016}
C.~Dong, C.~C. Loy, K.~He, and X.~Tang.
\newblock Image super-resolution using deep convolutional networks.
\newblock \emph{IEEE Trans.\ Pattern Anal.\ Mach.\ Intell.},
  \href{http://dx.doi.org/10.1109/TPAMI.2015.2439281}{38\penalty0 (2):\penalty0
  295--307}, 2016.

\bibitem[Dresdner et~al.(2022)Dresdner, Kochkov, Norgaard,
  Zepeda-N{\'u}{\~n}ez, Smith, Brenner, and Hoyer]{Dresdner_al_2022}
G.~Dresdner, D.~Kochkov, P.~Norgaard, L.~Zepeda-N{\'u}{\~n}ez, J.~A. Smith,
  M.~P. Brenner, and S.~Hoyer.
\newblock Learning to correct spectral methods for simulating turbulent flows,
  2022.
\newblock \doi{10.48550/ARXIV.2207.00556}.

\bibitem[Evensen(1994)]{Evensen_1994}
G.~Evensen.
\newblock Sequential data assimilation with a nonlinear quasi-geostrophic model
  using {M}onte {C}arlo methods to forecast error statistics.
\newblock \emph{J.\ Geophys.\ Res.},
  \href{http://dx.doi.org/10.1029/94JC00572}{99\penalty0 (C5):\penalty0
  10143--10162}, 1994.

\bibitem[Evensen(2007)]{Evensen_2007}
G.~Evensen.
\newblock \emph{Data Assimilation -- The Ensemble Kalman Filter}.
\newblock Springer, 1 edition, 2007.
\newblock \doi{10.1007/978-3-540-38301-7}.

\bibitem[Frerix et~al.(2021)Frerix, Kochkov, Smith, Cremers, Brenner, and
  Hoyer]{Frerix_al_2021}
T.~Frerix, D.~Kochkov, J.~Smith, D.~Cremers, M.~Brenner, and S.~Hoyer.
\newblock Variational data assimilation with a learned inverse observation
  operator.
\newblock In Ma. Meila and T.~Zhang, editors, \emph{Proceedings of the 38th
  International Conference on Machine Learning}, volume 139 of
  \emph{Proceedings of Machine Learning Research}, pages 3449--3458. PMLR,
  2021.

\bibitem[Fukami and Taira(2024)]{Fukami_Taira_2024}
K.~Fukami and K.~Taira.
\newblock Single-snapshot machine learning for super-resolution of turbulence.
\newblock \emph{J.\ Fluid Mech.},
  \href{http://dx.doi.org/10.1017/jfm.2024.1136}{1001:\penalty0 A32}, 2024.

\bibitem[Fukami et~al.(2019)Fukami, Fukagata, and
  Taira]{Fukami_Fukagata_Taira_2019}
K.~Fukami, K.~Fukagata, and K.~Taira.
\newblock Super-resolution reconstruction of turbulent flows with machine
  learning.
\newblock \emph{J.\ Fluid Mech.},
  \href{http://dx.doi.org/10.1017/jfm.2019.238}{870:\penalty0 106–120}, 2019.

\bibitem[Fukami et~al.(2021)Fukami, Fukagata, and
  Taira]{Fukami_Fukagata_Taira_2021}
K.~Fukami, K.~Fukagata, and K.~Taira.
\newblock Machine-learning-based spatio-temporal super resolution
  reconstruction of turbulent flows.
\newblock \emph{J.\ Fluid Mech.},
  \href{http://dx.doi.org/10.1017/jfm.2020.948}{909:\penalty0 A9}, 2021.

\bibitem[Fukami et~al.(2023)Fukami, Fukagata, and
  Taira]{Fukami_Fukagata_Taira_2023}
K.~Fukami, K.~Fukagata, and K.~Taira.
\newblock Super-resolution analysis via machine learning: a survey for fluid
  flows.
\newblock \emph{Theor.\ Comput.\ Fluid Dyn.},
  \href{http://dx.doi.org/10.1007/s00162-023-00663-0}{37:\penalty0 421--444},
  2023.

\bibitem[Gronskis et~al.(2013)Gronskis, Heitz, and
  M{\'e}min]{Gronskis_Heitz_Memin_2013}
A.~Gronskis, D.~Heitz, and E.~M{\'e}min.
\newblock Inflow and initial conditions for direct numerical simulation based
  on adjoint data assimilation.
\newblock \emph{J.\ Comp. Phys.},
  \href{http://dx.doi.org/10.1016/j.jcp.2013.01.051}{242:\penalty0 480--497},
  2013.

\bibitem[Hayase(2015)]{Hayase_2015}
T.~Hayase.
\newblock Numerical simulation of real-world flows.
\newblock \emph{Fluid Dynamics Research},
  \href{http://dx.doi.org/10.1088/0169-5983/47/5/051201}{47\penalty0
  (5):\penalty0 051201}, 2015.

\bibitem[He et~al.(2016)He, Zhang, Ren, and Sun]{He_Zhang_Ren_Sun_2016}
K.~He, X.~Zhang, S.~Ren, and J.~Sun.
\newblock Deep residual learning for image recognition.
\newblock In \emph{Proceedings of the IEEE Conference on Computer Vision and
  Pattern Recognition (CVPR)}, pages 770--778, 2016.

\bibitem[Hendrycks and Gimpel(2016)]{Hendrycks_Gimpel_2016}
D.~Hendrycks and K.~Gimpel.
\newblock Gaussian error linear units ({G}{E}{L}{U}{S}).
\newblock \emph{CoRR},
  \href{http://dx.doi.org/10.48550/arXiv.1606.08415}{abs/1606.08415}, 2016.

\bibitem[H{\oe}pffner et~al.(2005)H{\oe}pffner, Chevalier, Bewley, and
  Henningson]{Hoepffner_Chevalier_Bewley_Henningson_2005}
J.~H{\oe}pffner, M.~Chevalier, T.~R. Bewley, and D.~S. Henningson.
\newblock State estimation in wall-bounded flow systems. part 1. perturbed
  laminar flows.
\newblock \emph{J.\ Fluid Mech.},
  \href{http://dx.doi.org/10.1017/S0022112005004210}{534:\penalty0 263–294},
  2005.

\bibitem[Hunt et~al.(1988)Hunt, Wray, and Moin]{Hunt_al_1988}
J.C.R. Hunt, A.A. Wray, and P.~Moin.
\newblock Eddies, streams, and convergence zones in turbulent flows.
\newblock \emph{CTR Proc. Summer Program 1988}, pages 193--208, 1988.

\bibitem[Inubushi and Caulfield(2025)]{Inubushi_Caulfield_2025}
M.~Inubushi and C.~P. Caulfield.
\newblock Dimensional dependence of synchronisation in turbulence: insights
  from data assimilation and lyapunov analysis, 2025.
\newblock \doi{10.48550/arXiv.2508.16920}.

\bibitem[Kalman(1960)]{Kalman_1960}
R.~E. Kalman.
\newblock A new approach to linear filtering and prediction problems.
\newblock \emph{J.\ Basic Eng.},
  \href{http://dx.doi.org/10.1115/1.3662552}{82\penalty0 (1):\penalty0 35--45},
  1960.

\bibitem[Kalnay(2002)]{Kalnay_2002}
E.~Kalnay.
\newblock \emph{Atmospheric Modeling, Data Assimilation and Predictability}.
\newblock Cambridge University Press, 2 edition, 2002.
\newblock \doi{10.1017/CBO9780511802270}.

\bibitem[Kato and Obayashi(2013)]{Kato_Obayashi_2013}
H.~Kato and S.~Obayashi.
\newblock Approach for uncertainty of turbulence modeling based on data
  assimilation technique.
\newblock \emph{Comp.\ \& Fluids},
  \href{http://dx.doi.org/10.1016/j.compfluid.2012.09.002}{85:\penalty0 2--7},
  2013.

\bibitem[Kato et~al.(2015)Kato, Yoshizawa, Ueno, and
  Obayashi]{Kato_Yoshizawa_Ueno_Obayashi_2015}
H.~Kato, A.~Yoshizawa, G.~Ueno, and S.~Obayashi.
\newblock A data assimilation methodology for reconstructing turbulent flows
  around aircraft.
\newblock \emph{J.\ Comp. Phys.},
  \href{http://dx.doi.org/10.1016/j.jcp.2014.12.013}{283:\penalty0 559--581},
  2015.

\bibitem[Kelshaw et~al.(2022)Kelshaw, Rigas, and
  Magri]{Kelshaw_Rigas_Magri_2022}
D.~Kelshaw, G.~Rigas, and L.~Magri.
\newblock Physics-informed {C}{N}{N}s for super-resolution of sparse
  observations on dynamical systems.
\newblock In \emph{NeurIPS 2022, Machine Learning and the Physical Sciences
  Workshop}, 2022.

\bibitem[Kim et~al.(2021)Kim, Kim, Won, and Lee]{Kim_Kim_Won_Lee_2021}
H.~Kim, J.~Kim, S.~Won, and C.~Lee.
\newblock Unsupervised deep learning for super-resolution reconstruction of
  turbulence.
\newblock \emph{J.\ Fluid Mech.},
  \href{http://dx.doi.org/10.1017/jfm.2020.1028}{910:\penalty0 A29}, 2021.

\bibitem[Kim et~al.(1987)Kim, Moin, and Moser]{Kim_Moin_Moser_1987}
J.~Kim, P.~Moin, and R.~Moser.
\newblock Turbulence statistics in fully developed channel flow at low
  {R}eynolds number.
\newblock \emph{J.\ Fluid Mech.},
  \href{http://dx.doi.org/10.1017/S0022112087000892}{177:\penalty0 133--166},
  1987.

\bibitem[Kingma and Ba(2015)]{Kingma_Ba_2015}
D.~P. Kingma and J.~L. Ba.
\newblock Adam: a method for stochastic optimization.
\newblock In Y.~Bengio and Y.~LeCun, editors, \emph{3rd International
  Conference on Learning Representations, ICLR 2015, San Diego, CA, USA, May
  7-9, 2015, Conference Track Proceedings}, 2015.
\newblock \doi{10.48550/arXiv.1412.6980}.

\bibitem[Kochkov et~al.(2021)Kochkov, Smith, Alieva, Wang, Brenner, and
  Hoyer]{Kochkov_al_2021}
D.~Kochkov, J.~A. Smith, A.~Alieva, Q.~Wang, M.~P. Brenner, and S.~Hoyer.
\newblock Machine learning–accelerated computational fluid dynamics.
\newblock \emph{Proc. Natl. Acad. Sci.},
  \href{http://dx.doi.org/10.1073/pnas.2101784118}{118\penalty0 (21):\penalty0
  e2101784118}, 2021.

\bibitem[Lalescu et~al.(2013)Lalescu, Meneveau, and
  Eyink]{Lalescu_Meneveau_Eyink_2013}
C.~C. Lalescu, C.~Meneveau, and G.~L. Eyink.
\newblock Synchronization of chaos in fully developed turbulence.
\newblock \emph{Phys. Rev. Lett.},
  \href{http://dx.doi.org/10.1103/PhysRevLett.110.084102}{110:\penalty0
  084102}, 2013.

\bibitem[Li et~al.(2020)Li, Zhang, Dong, and
  Abdullah]{Li_Zhang_Dong_Abdullah_2020}
Y.~Li, J.~Zhang, G.~Dong, and N.~S. Abdullah.
\newblock Small-scale reconstruction in three-dimensional kolmogorov flows
  using four-dimensional variational data assimilation.
\newblock \emph{J.\ Fluid Mech.},
  \href{http://dx.doi.org/10.1017/jfm.2019.960}{885:\penalty0 A9}, 2020.

\bibitem[Linkmann(2018)]{Linkmann2018}
M.~Linkmann.
\newblock Effects of helicity on dissipation in homogeneous box turbulence.
\newblock \emph{J. Fluid Mech.},
  \href{http://dx.doi.org/10.1017/jfm.2018.709}{856:\penalty0 79–102}, 2018.

\bibitem[McKay et~al.(2017)McKay, Linkmann, Clark, Chalupa, and
  Berera]{McKay_Linkmann_Clark_Chalupa_Berera_2017}
M.~E. McKay, M.~Linkmann, D.~Clark, A.~A. Chalupa, and A.~Berera.
\newblock Comparison of forcing functions in magnetohydrodynamics.
\newblock \emph{Phys. Rev. Fluids},
  \href{http://dx.doi.org/10.1103/PhysRevFluids.2.114604}{2:\penalty0 114604},
  2017.

\bibitem[Mo and Magri(2025)]{Mo_Magri_2025}
Y.~Mo and L.~Magri.
\newblock Reconstruction of three-dimensional turbulent flows from sparse and
  noisy planar measurements: A weight-sharing neural network approach, 2025.
\newblock \doi{10.48550/arXiv.2509.18687}.

\bibitem[Mons et~al.(2017)Mons, Chassaing, and
  Sagaut]{Mons_Chassaing_Sagaut_2017}
V.~Mons, J.-C. Chassaing, and P.~Sagaut.
\newblock Optimal sensor placement for variational data assimilation of
  unsteady flows past a rotationally oscillating cylinder.
\newblock \emph{J.\ Fluid Mech.},
  \href{http://dx.doi.org/10.1017/jfm.2017.313}{823:\penalty0 230–277}, 2017.

\bibitem[Page(2025{\natexlab{a}})]{Page2025rev}
J.~Page.
\newblock Computation of simple invariant solutions in fluid turbulence with
  the aid of deep learning.
\newblock \emph{Nonlinear Dyn.}, 2025{\natexlab{a}}.
\newblock ISSN 1573-269X.

\bibitem[Page(2025{\natexlab{b}})]{Page_2025}
J.~Page.
\newblock Super-resolution of turbulence with dynamics in the loss.
\newblock \emph{J.\ Fluid Mech.},
  \href{http://dx.doi.org/10.1017/jfm.2024.1202}{1002:\penalty0 R3},
  2025{\natexlab{b}}.

\bibitem[Page et~al.(2024)Page, Norgaard, Brenner, and
  Kerswell]{page2022recurrent}
J.~Page, P.~Norgaard, M.~P. Brenner, and R.~R. Kerswell.
\newblock Recurrent flow patterns as a basis for two-dimensional turbulence:
  Predicting statistics from structures.
\newblock \emph{Proc. Natl. Acad. Sci.},
  \href{http://dx.doi.org/10.1073/pnas.2320007121}{121\penalty0 (23):\penalty0
  e2320007121}, 2024.

\bibitem[Pope(2000)]{Pope_2000}
S.~B. Pope.
\newblock \emph{Turbulent Flows}.
\newblock Cambridge University Press, 5th edition, 2000.
\newblock \doi{10.1017/CBO9780511840531}.

\bibitem[Taira et~al.(2025)Taira, Rigas, and Fukami]{Taira_Rigas_Fukami_2025}
K.~Taira, G.~Rigas, and K.~Fukami.
\newblock Machine learning in fluid dynamics: A critical assessment, 2025.
\newblock \doi{10.48550/arXiv.2508.13430}.

\bibitem[Talagrand(1997)]{Talagrand_1997}
O.~Talagrand.
\newblock Assimilation of observations, an introduction.
\newblock \emph{J.\ Met.\ Soc.\ Japan. Ser. II},
  \href{http://dx.doi.org/10.2151/jmsj1965.75.1B_191}{75\penalty0
  (1B):\penalty0 191--209}, 1997.

\bibitem[Talagrand and Courtier(1987)]{Talagrand_Courtier_1987}
O.~Talagrand and P.~Courtier.
\newblock Variational assimilation of meteorological observations with the
  adjoint vorticity equation. {I}: Theory.
\newblock \emph{Q.\ J.\ Roy.\ Meteorol.\ Soc.},
  \href{http://dx.doi.org/10.1002/qj.49711347812}{113\penalty0 (478):\penalty0
  1311--1328}, 1987.

\bibitem[Um et~al.(2020)Um, Brand, Fei, Holl, and Thuerey]{Um_al_2020}
K.~Um, R.~Brand, Y.~R. Fei, P.~Holl, and N.~Thuerey.
\newblock Solver-in-the-loop: Learning from differentiable physics to interact
  with iterative {P}{D}{E}-solvers.
\newblock In H.~Larochelle, M.~Ranzato, R.~Hadsell, M.F. Balcan, and H.~Lin,
  editors, \emph{Adv.\ Neur.\ Inform.\ Process.\ Sys.}, volume~33, pages
  6111--6122. Curran Associates, Inc., 2020.

\bibitem[Vela-Mart{\'i}n(2021)]{VelaMartin_2021}
A.~Vela-Mart{\'i}n.
\newblock The synchronisation of intense vorticity in isotropic turbulence.
\newblock \emph{J.\ Fluid Mech.},
  \href{http://dx.doi.org/10.1017/jfm.2021.153}{913:\penalty0 R8}, 2021.

\bibitem[Wang and Zaki(2021)]{Wang_Zaki_2021}
M.~Wang and T.~A. Zaki.
\newblock State estimation in turbulent channel flow from limited observations.
\newblock \emph{J.\ Fluid Mech.},
  \href{http://dx.doi.org/10.1017/jfm.2021.268}{917:\penalty0 A9}, 2021.

\bibitem[Wang and Zaki(2022)]{Wang_Zaki_2022}
M.~Wang and T.~A. Zaki.
\newblock Synchronization of turbulence in channel flow.
\newblock \emph{J.\ Fluid Mech.},
  \href{http://dx.doi.org/10.1017/jfm.2022.397}{943:\penalty0 A4}, 2022.

\bibitem[Wang and Zaki(2025)]{Wang_Zaki_2025}
M.~Wang and T.~A. Zaki.
\newblock Variational data assimilation in wall turbulence: from outer
  observations to wall stress and pressure.
\newblock \emph{J.\ Fluid Mech.},
  \href{http://dx.doi.org/10.1017/jfm.2025.132}{1008:\penalty0 A26}, 2025.

\bibitem[Wang et~al.(2019)Wang, Wang, and Zaki]{Wang_Wang_Zaki_2019}
M.~Wang, Q.~Wang, and T.~A. Zaki.
\newblock Discrete adjoint of fractional-step incompressible {N}avier-{S}tokes
  solver in curvilinear coordinates and application to data assimilation.
\newblock \emph{J.\ Comp. Phys.},
  \href{http://dx.doi.org/10.1016/j.jcp.2019.06.065}{396:\penalty0 427--450},
  2019.

\bibitem[Wang et~al.(2022)Wang, Wang, and Zaki]{Wang_Wang_Zaki_2022}
Q.~Wang, M.~Wang, and T.~A. Zaki.
\newblock What is observable from wall data in turbulent channel flow?
\newblock \emph{J.\ Fluid Mech.},
  \href{http://dx.doi.org/10.1017/jfm.2022.295}{941:\penalty0 A48}, 2022.

\bibitem[Yasuda and Onishi(2023)]{Yasuda_Onishi_2023}
Y.~Yasuda and R.~Onishi.
\newblock Spatio-temporal super-resolution data assimilation ({S}{R}{D}{A})
  utilizing deep neural networks with domain generalization.
\newblock \emph{J.\ Adv. Model.\ Earth Sys.},
  \href{http://dx.doi.org/10.1029/2023MS003658}{15\penalty0 (11):\penalty0
  e2023MS003658}, 2023.

\bibitem[Yoshida et~al.(2005)Yoshida, Yamaguchi, and
  Kaneda]{Yoshida_Yamaguchi_Kaneda_2005}
K.~Yoshida, J.~Yamaguchi, and Y.~Kaneda.
\newblock Regeneration of small eddies by data assimilation in turbulence.
\newblock \emph{Phys. Rev. Lett.},
  \href{http://dx.doi.org/10.1103/PhysRevLett.94.014501}{94:\penalty0 014501},
  2005.

\bibitem[Zaki(2025)]{Zaki_2025}
T.~A. Zaki.
\newblock Turbulence from an observer perspective.
\newblock \emph{Ann.\ Rev.\ Fluid Mech.},
  \href{http://dx.doi.org/10.1146/annurev-fluid-030424-114735}{57:\penalty0
  311--334}, 2025.

\end{thebibliography}
